\documentclass[conference]{IEEEtran}
\pdfoutput=1

\makeatletter
\def\@ACM@checkaffil{
    \if@ACM@instpresent\else
    \ClassWarningNoLine{\@classname}{No institution present for an affiliation}%
    \fi
    \if@ACM@citypresent\else
    \ClassWarningNoLine{\@classname}{No city present for an affiliation}%
    \fi
    \if@ACM@countrypresent\else
        \ClassWarningNoLine{\@classname}{No country present for an affiliation}%
    \fi
}
\makeatother

\AtBeginDocument{%
  \providecommand\BibTeX{{%
    Bib\TeX}}}

\usepackage{siunitx}
\usepackage{float} 

\usepackage{booktabs}   
\usepackage{subcaption} 

\usepackage{hyperref}
\usepackage{amsmath}
\usepackage[vlined,ruled,linesnumbered,noresetcount]{algorithm2e}
\usepackage{cleveref}
\usepackage[normalem]{ulem}
\usepackage{listings}
\usepackage[export]{adjustbox}
\usepackage{cite}
\usepackage{amsmath,amssymb,amsfonts}
\usepackage{algorithmic}
\usepackage{graphicx}
\usepackage{textcomp}
\usepackage{xcolor}
\usepackage{amsthm, amssymb}
\theoremstyle{definition}
\newtheorem{definition}{Definition}[section]
\newtheorem{theorem}{Theorem}[section]
\newtheorem{lemma}{Lemma}[section]
\def\BibTeX{{\rm B\kern-.05em{\sc i\kern-.025em b}\kern-.08em
    T\kern-.1667em\lower.7ex\hbox{E}\kern-.125emX}}

\newcommand{\rb}[1]{{\color{red} #1}\normalcolor}

\SetKwProg{myproc}{Procedure}{}{}
\SetKwBlock{Type}{type}{}

\DontPrintSemicolon

\let\oldnl\nl
\newcommand{\nonl}{\renewcommand{\nl}{\let\nl\oldnl}}

\makeatletter
\newcommand{\removelatexerror}{\let\@latex@error\@gobble}
\makeatother

\makeatletter
\lst@Key{countblanklines}{true}[t]%
{\lstKV@SetIf{#1}\lst@ifcountblanklines}

\lst@AddToHook{OnEmptyLine}{%
	\lst@ifnumberblanklines\else%
	\lst@ifcountblanklines\else%
	\advance\c@lstnumber-\@ne\relax%
	\fi%
	\fi}
\makeatother

\newcommand{\here}[1]{{\bf [[[#1]]]}}
\newcommand{\ignore}[1]{}
\newcommand{\remove}[1]{}

\newcommand{\y}[1]{{\color{black} #1}\normalcolor}
\newcommand{\yy}[1]{{\color{black} #1}\normalcolor}
\newcommand{\lc}[1]{{\here{Lef: #1}}}

\newcommand{\lred}[1]{{\color{black} #1}\normalcolor}

\newcommand{\MESSI}{\mbox{MESSI}}
\newcommand{\Fresh}{\mbox{FreSh}}
\newcommand{\FreSh}{\mbox{FreSh}}

\newcommand{\Refresh}{\mbox{Refresh}}
\newcommand{\ReFresh}{\mbox{Refresh}}

\newcommand{\MESSIenh}{\mbox{MESSI-enh}}

\newcommand{\Backoff}{\mbox{Backoff}}

\newcommand{\FreshSub}{\mbox{Subtree}}
\newcommand{\FreshSTD}{\mbox{Standard}}
\newcommand{\FreshTreeCopy}{\mbox{TreeCopy}}

\newcommand{\CAS}{\mbox{\textit{CAS}}}
\newcommand{\FAI}{\mbox{\textit{FAI}}}

\newcommand{\True}{\mbox{\texttt{True}}}
\newcommand{\False}{\mbox{\texttt{False}}}
\newcommand{\Boolean}{\mbox{boolean}}

\newcommand{\Insert}{\mbox{\sc Insert}}
\newcommand{\TreeInsert}{\mbox{\sc TreeInsert}}

\newcommand{\MBInsert}{\mbox{\sc MBInsert}}
\newcommand{\Put}{\mbox{\sc Put}}
\newcommand{\Transform}{\mbox{\sc Transform}}

\newcommand{\Traverse}{\mbox{\sc Traverse}}
\newcommand{\SplitLeaf}{\mbox{\sc SplitLeaf}}
\newcommand{\DeleteMin}{\mbox{\sc DeleteMin}}

\newcommand{\BufferCreation}{\mbox{\sc BufferCreation}}
\newcommand{\TreePopulation}{\mbox{\sc TreePopulation}}
\newcommand{\Prunning}{\mbox{\sc Prunning}}
\newcommand{\Refinement}{\mbox{\sc Refinement}}
\newcommand{\QueryAnswering}{\mbox{\sc QueryAnswering}}
\newcommand{\UpdateBSF}{\mbox{\sc UpdateBSF}}
\newcommand{\FindNode}{\mbox{\sc FindNode}}
\newcommand{\TotalNodes}{\mbox{\sc TotalNodes}}
\newcommand{\HelpTree}{\mbox{\sc HelpTree}}
\newcommand{\NextIndex}{\mbox{\sc NextIndex}}

\newcommand{\FI}{\mbox{\sf FI-Based}}
\newcommand{\FINoSum}{\mbox{\sf FI-Based-NoSum}}
\newcommand{\DoAllSplit}{\mbox{\sf DoAll-Split}}

\newcommand{\CASBased}{\mbox{\sf CAS-Based}}

\newcommand{\Search}{\mbox{Search}}

\newcommand{\BC}{\mbox{$\mathit{BC}$}}
\newcommand{\TP}{\mbox{$\mathit{TP}$}}
\newcommand{\PS}{\mbox{$\mathit{PS}$}}
\newcommand{\RS}{\mbox{$\mathit{RS}$}}
\newcommand{\RawData}{\mbox{$\mathit{RawData}$}}

\SetKw{Continue}{continue}
\SetKw{Break}{break}
\SetKw{WaitUntil}{wait until}
\SetKw{And}{and}
\SetKw{Or}{or}
\SetKw{Mod}{mod}
\SetKw{Allocate}{allocate}
\SetKw{Execute}{execute}
\SetKw{Using}{using}
\SetKw{Replacing}{replacing}
\SetKw{Adding}{adding}
\SetKw{Add}{add}
\SetKw{To}{to}
\SetKw{Each}{each}
\SetKw{Do}{do}
\SetKw{CASS}{CAS}
\SetKw{Extends}{extends}
\SetKw{New}{new}
\SetKw{With}{with}
\SetKw{Split}{split}
\SetKw{Distribute}{distribute}
\SetKw{Populate}{populate}
\SetKw{Not}{not}

\newcommand{\comnospace}{\mbox{$\triangleright$}}
\newcommand{\com}{\mbox{\comnospace\ }}

\begin{document}

\title{FreSh: A Lock-Free Data Series Index}

\author{
	\IEEEauthorblockN{Panagiota Fatourou}
	\IEEEauthorblockA{
		\textit{ICS-FORTH} \&\\
		\textit{University of Crete } \\
		faturu@csd.uoc.gr}
	\and
	\IEEEauthorblockN{Eleftherios Kosmas}
	\IEEEauthorblockA{
		\textit{Hellenic Mediterranean} \\
		\textit{University } \\
		ekosmas@hmu.gr}
	\and
	\IEEEauthorblockN{Themis Palpanas}
	\IEEEauthorblockA{
		\textit{LIPADE, Universit{\'e} Paris Cit{\'e} \&} \\
		\textit{French University Institute (IUF)} \\
		themis@mi.parisdescartes.fr}
	\and
	\IEEEauthorblockN{George Paterakis}
	\IEEEauthorblockA{
		\textit{ICS-FORTH} \& \\
		\textit{University of Crete } \\
		csdp1311@csd.uoc.gr}
}

\maketitle

\begin{abstract}
%
We present \FreSh, a {\em lock-free} data series index
that exhibits good performance (while being robust). 
\Fresh\ is based on \Refresh, which is 
a {\em generic approach} we have developed for supporting lock-freedom  
in an efficient way on top of any {\em locality-aware} data series index. 
We believe \Refresh\ is of independent interest and can be used to get
well-performed lock-free versions of other locality-aware blocking data structures.
For developing \FreSh, we first studied in depth the design decisions of current 
state-of-the-art data series indexes, and the principles governing their performance. 
This led to a theoretical framework, which 
enables the development and analysis of data series indexes in a modular way. 
The framework allowed us to apply \Refresh, repeatedly,
to get lock-free versions of the different phases of a family of data series indexes.
Experiments with several synthetic and real datasets 
illustrate that 
\FreSh\ achieves performance that is as good as that 
of the state-of-the-art \emph{blocking} in-memory data series index. This shows that the 
helping mechanisms of \Fresh\ are light-weight, respecting certain principles that are crucial for performance
in locality-aware data structures.This paper was published in SRDS 2023. 
\end{abstract}

\section{Introduction}
\label{sec:intro}

Processing big collections of data series is of paramount importance 
for a wide spectrum of applications, 
\yy{across many domains, such as: operation health monitoring in data centers, vehicles and
manufacturing processes, internet of things data analysis, environmental and climate monitoring,
energy consumption analysis, decision taking in financial markets, telecommunications traffic
analysis, detection of medical and health problems, improvement of web-search results, identification
of pests invading agricultural crops, etc.~\cite{DBLP:journals/sigmod/Palpanas15,DBLP:journals/dagstuhl-reports/BagnallCPZ19,Palpanas2019}. }
In the heart of 
analyzing such collections 
lies the process of similarity search. 
Given a 
query series $Q$, {\em similarity search} 
returns a set of data series from the collection that have the closest 
distance 
to $Q$. 
Similarity search 
comes at considerable cost, 
due to 
very large size of data series collections, and 
the
high dimensionality (i.e., length) of the data series that modern applications need to analyze. 
To address these challenges, current state-of-the-art data series 
indexes~\cite{DBLP:journals/pvldb/EchihabiZPB18,isax2plus,wang2013data,peng2018paris,parisplus,peng2020messi,PFP21-I,PFP21-II,hercules,dumpy}
are based on data series summarization. They develop a tree index containing
data series summaries used to prune the series collection in order to
restrict the execution of costly computations only to a small subset of it.


State-of-the-art data series indexes~\cite{DBLP:journals/pvldb/EchihabiZPB18,isaxfamily,peng2018paris,peng2020messi,PFP21-I,PFP21-II,hercules} 
exploit the parallelism supported
by modern multicore machines, 
but 
are largely lock-based to achieve synchronization. 
Using locks results in {\em blocking} implementations: if a thread that
holds a lock delays, other threads block, without making any progress,  until the lock is released. 
Such thread delays  can degrade performance. 
Some
applications (eg, operation health-monitoring in nuclear plants, or gravitational-wave detection in
astrophysics), are sensitive to delays, and would benefit from our approach.
Locks may also
result in known problems, such as deadlock, priority inversion, and lock convoying~\cite{F04}.

\remove{
\rb{
\sout{Lock-based programming has particular disadvantages (discussed and studied
extensively in publications in PODC, DISC, PPoPP, SPAA, etc.). It may result in problems such as
deadlocks, priority inversion, lock convoying, etc., and it is undesirable in asynchronous settings,
where the thread holding the lock could be slow, experience page faults, or be paused by the system
scheduler.} 
}
}

{\em Lock-freedom}~\cite{HS08} is a widely-studied property when designing concurrent
trees~\cite{EFRB10,EFHR14,FR2018,ABF+22} and other data structures~\cite{F04,HS08,FKR18}. 
It avoids the use of locks, ensuring that the system, as a whole, 
makes progress, 
\yy{independently of delays (or failures) of threads. 
The relative performance of lock-free vs. lock-based algorithms depends on the setting:
when threads are pinned to distinct cores, algorithms using fine-grained locks do well. However, in
oversubscribed settings (more threads than cores), lock-based algorithms can suffer due to threads
getting de-scheduled while holding a lock. Lock-free algorithms address these issues, but may be
complicated and result in worse performance in settings with no delays/failures.
Designing lock-free data series indexes, which exhibit good performance, is
the focus of this paper: we develop a lock-free data-series index, which always produces the correct output, 
while maintaining the good performance of state-of-the-art lock-based indexes.}
\remove{
\rb{
The relative performance of lock-free vs. lock-based algorithms depends on the setting:
when threads are pinned to distinct cores, algorithms using fine-grained locks do well. However, in
oversubscribed settings (more threads than cores), lock-based algorithms can suffer due to threads
getting de-scheduled while holding a lock. Lock-free algorithms address these issues, but may be
complicated and result in worse performance in settings with no delays/failures. One of our main
contributions is a lock-free data-series (DS) index that exhibits the same good performance as the
state-of-the-art lock-based approach.
}
}

\noindent
{\bf Challenges.} 
To achieve lock-freedom, some form of {\em helping} is usually employed. 
That is, appropriate mechanisms are provided to make threads aware 
of the work that other threads perform, 
so that a thread may help others to complete their work whenever needed. 
Unfortunately, conventional helping mechanisms
are rather expensive and often introduce high overheads~\cite{F04,HS08,7515610,Williams2012CCI}. 
For this reason, the vast majority of the software stack is still based on locks. 
Ensuring lock-freedom while maintaining the good performance of 
existing data series indexes is a major challenge. 

State-of-the-art 
indexes are designed \y{to (a) maintain some form of {\em data locality}, 
and (b) avoid synchronization as much as possible.} 
For instance, 
they often separate the data into {\em disjoint sets}, and have a distinct thread
manipulate the data of each set~\cite{parisplus,PFP21-I,PFP21-II}. 
This processing pattern 
enables threads to work in parallel and independently from each other,
resulting in reduced synchronization and communication costs. 
These principles for reduced communication and synchronization 
are easily achieved when locks (or barriers) are utilized~\cite{peng2018paris,peng2020messi,PFP21-I,PFP21-II}.
However, the way helping works in conventional lock-free data structures is inherently incompatible to these
principles, thus making it 
challenging to implement helping on top of such indexes without sacrificing  them.
%
Providing lock-freedom while maintaining load-balancing among threads, 
and ensuring good data pruning are further challenges 
to address. 

State-of-the-art data series indexes encompass several data processing phases, 
which often employ different data structures
to accomplish their efficient processing. 
Coming up with
lock-free versions of these data structures, while respecting the communication 
and synchronization cost principles that govern existing 
 indexes, is another major challenge to address.

In order to develop a {\em generalized approach} for supporting lock-freedom 
on top of data series indexes 
in a {\em systematic way}, we need to study and understand 
the design decisions of state-of-the-art
indexes and the performance principles that govern them. 
Then, we need appropriate abstractions for the data series processing stages and their properties, 
as well as a set of design principles that need to be respected for efficiency. 
\y{Accomplishing these goals leads to additional challenges.} 

\noindent
{\bf Our approach.}
We propose \Fresh, a novel {\em lock-free} data series index, that efficiently
addresses all of the above challenges. 
Our experimental analysis shows that 
the performance of \Fresh\ is as good
as that of \MESSI~\cite{PFP21-I}, the state-of-the-art 
concurrent data-series index, which is lock-based. 
This attests to the high efficiency of 
the helping scheme we propose for \Fresh. 
Moreover, in many cases, 
\Fresh\ performs better than \MESSI, 
as it allows
for increased parallelism when constructing the tree index.
Note that if threads {\em crash}, \MESSI\ 
(and all other lock-based approaches~\cite{peng2018paris,PFP21-I,PFP21-II,hercules})
never terminate (so, we do not provide experiments for this case).
\Fresh\ always successfully and correctly terminates.

To get \Fresh, we developed a generic approach, called \Refresh, which can be applied on top of
a family of state-of-the-art blocking indexes to provide lock-freedom
without adding any cost. 
\ReFresh\ introduces the concept
of {\em locality-aware lock-freedom} which encompasses
the properties of data locality, high parallelism, low synchronization cost,
and load balancing met in the designs of many existing parallel data series indexes.
None of the
conventional lock-free techniques we are aware of has been designed with the goal of respecting these properties.
Indeed, our experiments show that such conventional techniques 
result in significantly lower performance. 

\ReFresh\ respects the workload and data separation of the underlying data series index, 
in order to not hurt the degrees of parallelism and load balancing of the index.
Moreover, it provides a mechanism for threads to {\em determine}
whether a specific part of a workload has been processed, and {\em help} only whenever necessary.
\Refresh\ introduces two modes of execution for each thread: 
(i) {\em expeditive} and (ii) {\em standard}.
A thread executing in {\em standard} mode may incur synchronization overhead,
as it needs to synchronize with helper threads; a thread executing in {expeditive} mode executes a code that avoids \y{ synchronization altogether}. 
A thread starts by processing its assigned workload in expeditive mode. 
Helping is performed only after a thread has finished processing its own workload.
Then, threads have to synchronize to execute on standard mode. 
This way, \Refresh\ maintains the synchronization and communication costs as low as that of the underlying index.

\Refresh\ can be applied on top of any {\em locality-aware algorithm} 
(Section~\ref{sec:llf}) to get a lock-free version of it. 
\Fresh\ (Section~\ref{sec:fresh}) follows 
the design decisions of locality-aware iSAX-based indexes~\cite{isaxfamily,PFP21-I} (see Section~\ref{sec:prelim}). 
However, to develop \Fresh, we had to replace all data structures 
of the original index~\cite{PFP21-I} with corresponding locality-aware lock-free versions;
we present lock-free implementations of several concurrent data structures, such as  
binary trees and priority queues (Section~\ref{sec:fresh}).
The proposed lock-free tree contains several new ideas. 
Previous solutions~\cite{EFRB10,DBLP:conf/podc/EllenFHR13,FR2018,NRM20} require that when a key is inserted in a leaf $\mathit{\ell}$, 
$\mathit{\ell}$ is copied and updated locally, and then
replaced in the shared tree. This results in bad performance. 
Instead, we designed a novel algorithm for updating leaves, 
which provides enhanced parallelism compared to existing algorithms.
Another novelty 
is the support of the new implementations for the expeditive and standard execution modes,
which was a challenge by itself, as synchronisation is
needed to transfer from one mode to the other.
We believe that these implementations, as well as \Refresh, 
could be employed to get highly-efficient lock-free versions of 
several other big data-processing solutions.

To be able to apply \ReFresh\ in a systematic way throughout all processing stages of an iSAX-based index,
we introduce the abstraction of a {\em traverse object} (Section~\ref{sec:traverse}).
The traverse object is an abstract data type that leads to a generic methodology for designing an iSAX-based index in a modular way.
It abstracts the main processing pattern used during the operation of
iSAX-based indexes. 
Specifically, the iSAX-based index can be implemented via traverse object operations.
\yy{The introduction of the traverse object is one of the novelties of our work. }

\noindent{\bf Contributions.} 
Our contributions are summarized as follows.

\noindent{$\bullet$ We develop a theoretical framework 
for supporting lock-freedom in a systematic way on top of highly-efficient data series indexes.
In particular, we present \Refresh, a novel {\em generic approach} that can be applied on top of any locality-aware data series algorithm
to ensure lock-freedom.} 

\noindent{$\bullet$ Based on \Refresh, we develop \Fresh, the first {\em lock-free}, efficient iSAX-based data series index. 
To get \Fresh, we present new lock-free implementations of several data structures 
which support the needed functionality.}

\noindent{$\bullet$ Our experiments, with large synthetic and real datasets, demonstrate that \Fresh\ performs as good as 
the state-of-the-art {\em blocking} 
index, thus paying no penalty for providing lock-freedom (and in many cases achieves better performance). 

\noindent{$\bullet$ Experiments show that by providing lock-freedom without jeopardizing locality-awareness,
\Fresh\ outperforms by far several lock-free baselines we have designed, 
based on conventional approaches for ensuring lock-freedom. }

\noindent{$\bullet$ We present a theoretical framework, which introduces the traverse object,
and utilizes it to enable the development 
of locality-aware data series indexes
in a modular way.}

\remove{
\rb{
We will follow the reviewer’s advice and add figures with an overview of the proposed approach, as
well as illustrations/intuitions of specific points (iSAX-based indexing and their implementation from
traverse objects, \Refresh\ functioning, \FreSh\ design). We will correct typos and address minor
comments.
}
}

\section{Preliminaries and Related Work}
\label{sec:prelim}

\noindent
{\bf Data Series, Indexing, and Similarity Search.} 
A {\em data series} (DS) of {\em size} (or {\em dimensionality}) $n$ 
is a sequence of $n$ (value, position) pairs. 
%
The {\em Piecewise Aggregate Approximation (PAA)}~\cite{DBLP:journals/kais/KeoghCPM01} of a data series 
is a vector of $w$ components which are calculated by splitting the x-axis  
into $w$ equal segments, and representing each segment with the mean value of the corresponding points (depicted by the black horizontal lines in Figure~\ref{fig:from_ds_to_iSAX}(b)). 
To calculate the iSAX summary~\cite{shieh2008sax} of the data series, 
the y axis is partitioned into a number 
of regions 
and a bit representation is introduced for each region. 
The {\em iSAX summary} is a vector of $w$ components that represent each of the $w$ segments of the series
not by the real value of the PAA, but 
with the symbol of the region the PAA falls into, forming the word $10_2 00_2 11_2$ shown in Figure~\ref{fig:from_ds_to_iSAX}(c) (subscripts denote the number of bits used to represent the symbol of each segment). 
The number of bits can be different for each region, and this enables 
the creation of a hierarchical tree index ({\em iSAX-based tree index}),
as shown in Figure~\ref{fig:from_ds_to_iSAX}(d). 
The index is implemented as a leaf-oriented tree with each leave 
storing up to $M$ keys. 
During an insertion, if the appropriate leaf $\ell$ has room, 
the new key is placed in $\ell$. 
Otherwise, $\ell$ is {\em split}: it is replaced by a subtree consisting of an internal node 
and two leaves that receive the keys of $\ell$. 
If one of the newly created leaves is empty, the splitting process is repeated.
For more details on iSAX-based indices, see~\cite{isaxfamily}.


\begin{figure}
	\centering
	\begin{subfigure} [b]{0.13\textwidth}
		\centering
		\includegraphics[width=\textwidth]{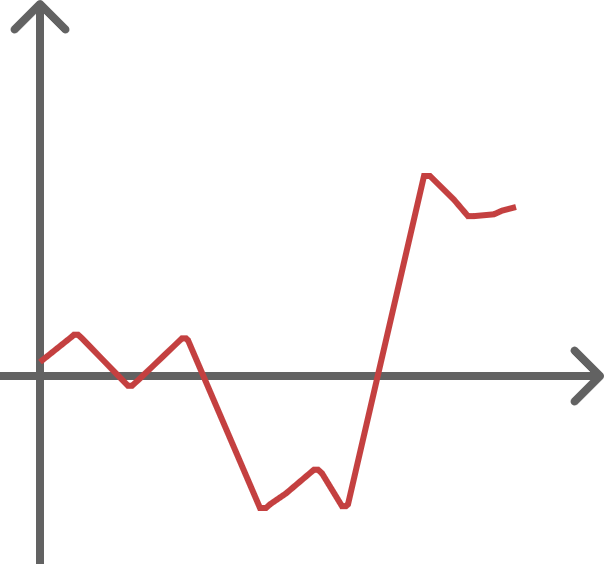}
		\caption{Data Series}
	\end{subfigure}
	\begin{subfigure} [b]{0.13\textwidth}
		\centering
		\includegraphics[width=\textwidth]{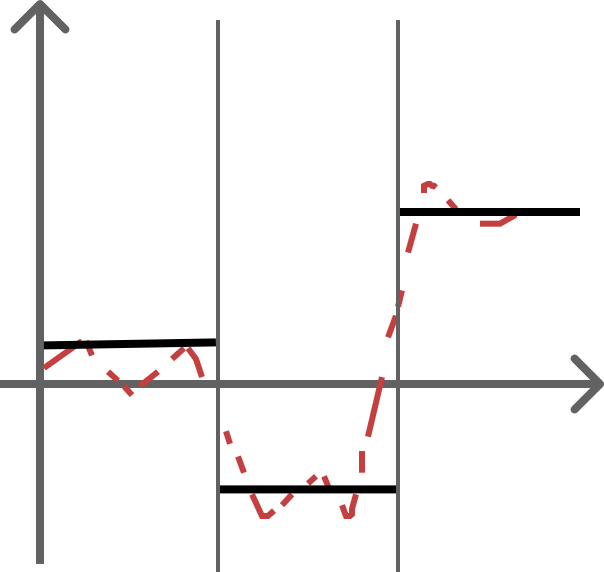}
		\caption{PAA Summary}
		\label{}
	\end{subfigure}
	\begin{subfigure} [b]{0.177\textwidth}
		\centering
		\includegraphics[width=\textwidth]{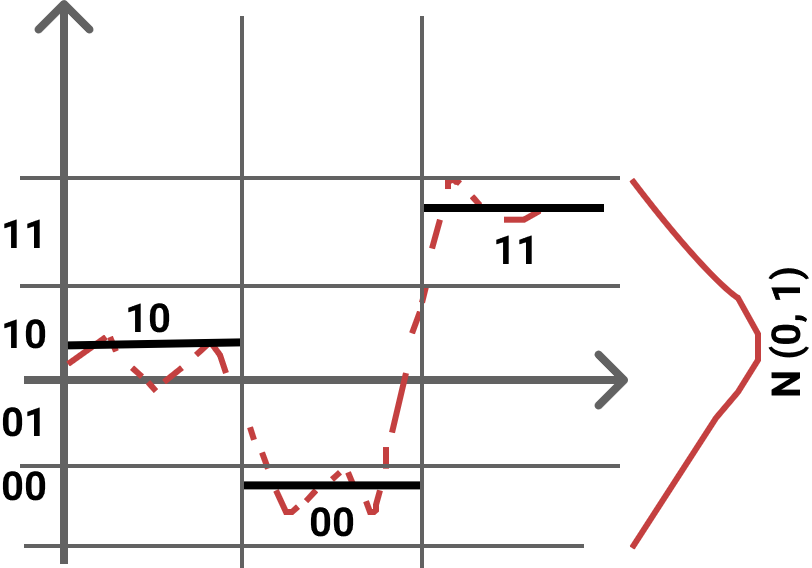}
		\caption{iSAX Summary}
		\label{}
	\end{subfigure}
	\begin{subfigure} [b]{0.38\textwidth}
		\centering
		\includegraphics[width=\textwidth]{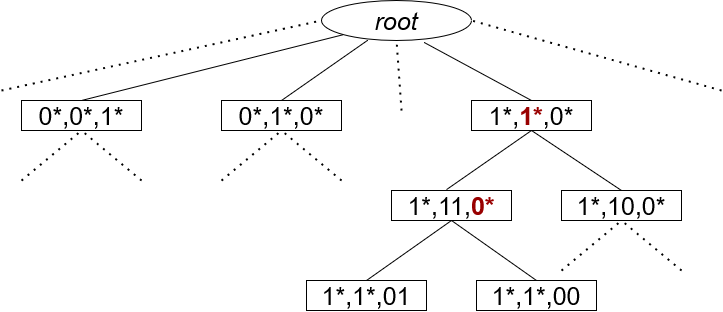}
		\caption{iSAX Tree}
		\label{}
	\end{subfigure}
	\vspace*{-0.2cm}
	\caption{From data series to iSAX index}
	\vspace*{-0.5cm}
	\label{fig:from_ds_to_iSAX}
\end{figure} 

We focus on {\em exact similarity search} (a.k.a. exact {\em 1-NN})
which returns the data series from a collection that is 
the most similar to a query data series. Similarity is measured
based on Euclidean Distance (ED), 
but our techniques are general enough to work for
other popular {\em similarity measures}, such as Dynamic Time Warping (DTW)~\cite{rakthanmanon2012searching}. 
%
We call
the distance between 
the {\em iSAX summaries} of two data series {\em lower-bound distance}. 
The way this distance is calculated ensures the {\em pruning property}: 
The lower bound distance between two data series is always smaller than or equal to their euclidean distance, 
which we call {\em real distance}. This
property enables pruning of data series during query answering:
A data series can be {\em pruned}, if its lower bound distance from the query series $Q$
exceeds any collection series' real distance to $Q$.

\noindent

\noindent\textbf{iSAX-Based Indexing.} 
Concurrent iSAX-based indexes \cite{peng2018paris,parisplus,peng2020messi,PFP21-I,PFP21-II} 
work in two phases. 
During the {\em tree index construction phase} (1st phase), a set of {\em worker threads} work on a collection 
of input data series (i.e., {\em raw data}), calculate an iSAX summary 
for each one of them, and build a {\em tree index} containing pairs of
iSAX summaries and pointers to the corresponding data series. 
In an iSAX-based index, these pairs are first stored 
into a set of array buffers, i.e., {\em summarization buffers} ({\em buffers creation} stage).
Then, the worker threads traverse these buffers and insert their entries in the tree index ({\em tree population} stage).
Data series that have similar summarizations are placed into the same buffer and later in the same root subtree of the index tree. 
This ensures high parallelism, a good degree of locality, 
and low synchronization overhead in building the index tree.  

Given a query data series $Q$, the following actions occur during {\em query answering}. 
A thread calculates the iSAX summary of $Q$ and uses it to traverse
a path of the tree index, reaching a leaf $\ell$.
Then, the thread calculates the {\em real distance} between $Q$ and each of the data series 
of $\ell$, and stores the smallest distance among them
in a variable called {\em BSF}.
This distance serves as an initial approximate query answer.
Query answering proceeds in two stages. A set of {\em query answering threads} traverse the tree and use BSF 
%
to select those data series that are potential {\em candidate series} for being the final answer to $Q$
({\em pruning 
\lred{stage}}). 
Those nodes whose lower bound distance to $Q$ is larger than BSF are {\em pruned}. 
%
The candidate series are often stored in (one or more) priority queues~\cite{parisplus,PFP21-I,PFP21-II}. 
Multiple threads process the elements of the priority queues
by calculating their real distances from $Q$ ({\em refinement stage}), 
and updating the BSF each time a new minimum is met.
At the end of the query answering phase, the final answer is contained in BSF. 
{\em Barriers} among threads are often used at the end of each stage 
to ensure correctness. Locks synchronize threads 
when they accesss the same parts of data structures. 

\noindent{\bf System.} 
We consider a shared-memory system of $N$ threads which are executed 
asynchronously and communicate by accessing shared objects. A shared object $O$ can be atomically
read or written. 
Moreover, \FAI(O,v) atomically reads the current value of $O$,
adds the value $v$ to it and returns the value read. A \CAS($O, u, v$) reads the value of $O$ 
and if it is equal to $u$, it changes it to $v$ and returns \True; otherwise, 
$O$ remains unchanged and \False\ is returned.

\remove{
Threads may fail by either crashing, or by experiencing some delay. When a thread {\em crashes},
it stops executing its algorithm and never recovers. When a thread experiences a delay,
it is temporarily unavailable, but it later resumes its execution. 
For example, a core may become slow 
(thus, causing delays to the threads executing on it) 
due to power consumption issues, or to overheating~\cite{inteloverheating}. 
}

Threads may experience delays (e.g., due to page faults, power consumption issues or overheating~\cite{inteloverheating})
or they may fail by crashing (e.g., due to software errors). 
An algorithm is {\em blocking} if a thread has to wait for actions to be taken by other threads in order to make progress.
{\em Lock-freedom} guarantees that the system as a whole continues to make progress, 
independently of the speed of threads or their failures.

\subsection{Other Related Work}

Several tree-based techniques for efficient and scalable data series similarity search 
have been proposed in the literature~\cite{DBLP:journals/pvldb/EchihabiZPB18,DBLP:journals/pvldb/EchihabiZPB19,DBLP:conf/edbt/EchihabiZP21,DBLP:journals/pvldb/EchihabiPZ21}, including approximate~\cite{DBLP:journals/pvldb/AziziEP23,DBLP:journals/kais/LevchenkoKYAMPS21} and progressive~\cite{DBLP:conf/sigmod/GogolouTEBP20,DBLP:journals/tvcg/JoSF20,DBLP:conf/sigmod/LiZAH20,DBLP:journals/vldb/EchihabiTGBP23} solutions. 
Out of those, the iSAX-based indexes~\cite{isaxfamily} have proven to be very competitive in terms of both index building and query answering time performance~\cite{DBLP:journals/pvldb/EchihabiZPB18,DBLP:journals/pvldb/EchihabiZPB19,hercules,odyssey,dumpy}.
These indexes also include parallel and distributed solutions 
that 
make use of modern hardware (e.g., SIMD, multi-core, multi-socket, GPU), such as ParIS+~\cite{parisplus}, MESSI~\cite{PFP21-I}, and SING~\cite{PFP21-II}, as well as distributed computation, 
such as DPiSAX~\cite{dpisax,dpisaxjournal} and Odyssey~\cite{odyssey}. 

The first lock-free implementation of a concurrent search tree appears in~\cite{EFRB10}. 
We use the main ideas from that paper to come up with a baseline algorithm, which we 
discuss and experimentally compare with \Fresh\ in Section~\ref{section:evaluation}. 
Many other non-blocking concurrent search trees have appeared in the literature (e.g.,~\cite{BER14,HL16,ABF20,HJ12,NRM20,CNT14,BP12,EFHR14,FR2018,ABF+22}. 
The novelty of the tree implementation we present in Section~\ref{sec:fresh}
is that it allows multiple insert operations to concurrently update (in a lock-free way) the array that stores the data
in a (fat) leaf. 
Additionally, it supports the expeditive-standard mode of execution. 
These innovations result in enhanced parallelism and better performance. 
Our algorithm is designed to only provide the functionality needed to implement traverse objects.
The above-mentioned implementations support different functionalities, have different goals, or
have been designed for other settings.

Concurrent priority queues appear in~\cite{AK15-I,RT21,WG15,SUNDELL2005609,tamir_et_al,LJ13}. 
In the baseline lock-free implementations we developed, 
we use a skip-list based lock-free priority queue~\cite{LJ13}, which has been shown to perform well. 
Our experiments show that the scheme of priority queues we designed for \Fresh, 
outperforms by far this implementation (Section~\ref{section:evaluation}). 

\remove{
Universal constructions~\cite{FK11spaa,FK12ppopp,FK14,FK17opodis,FK09,FK20,FKK18,EF+16,FKK22} can be used to provide wait-free or non-blocking
concurrent versions of any sequential data structure.
However, because of their generality, they are usually
less efficient than implementations tailor-made for specific sequential data
structures. 
The algorithms in~\cite{FK11spaa,FK14,FKK22} are highly efficient 
for implementing shared objects of small size (such as stacks and queues), but they are not appropriate
for our purpose.
}

The idea of transforming an algorithm to get an implementation
that ensures a different progress guarantee is not new.
Examples of such transformations appear in~\cite{SP14,ELM05,GKK06}
but they have all been introduced to solve different problems
and
%
the main technique of \Refresh\ departs from
all these approaches.

\section{Traverse Objects}
\label{sec:traverse}

\remove{The traverse object has the power to
abstract the main processing pattern met during the operation of
iSAX-based indexes (and possibly also of other big-data processing applications).  
Specifically, the iSAX-based index can be implemented
as a sequence of operations on traverse objects. 
This way, we provide a generic and formal methodology for designing iSAX-based indexes. 
Specifically, the design of an iSAX-based index largerly decomposes to determining 
and implementing a number of traverse objects. 
This results in a structural construction of an iSAX-based index
whose properties (correctness and/or progress) can be studied and anlyzed more easily
than in a flat index design. 

We first provide a definition of a traverse
object, and then we present how an iSAX-based index can be implemented
as a sequence of operations on traverse objects. 

}
Each of the last three stages of an iSAX-based index processes 
data that are produced by the previous stage. 
The first stage processes the original collection of data.
This processing pattern has inspired the definition of the traverse object,
\yy{whose sequential specification is provided below.}

\begin{definition}
\label{def:traverse}
Let $U$ be a universe of elements. 
A {\em traverse object} $S$ stores elements of $U$
(not necessarily distinct) and supports the following operations:
\begin{itemize}
\item {\em \Put($\mathit{S,e,param}$)}, which adds an element $e \in U$ in $S$;
$\mathit{param}$  is an optional argument that allows an implementation
to pass certain parameters in \Put. 
\item {\em \Traverse($\mathit{S,f,param,del}$)}, which traverses $S$ 
and applies the function $f$ 
on each of the traversed elements. 
If the $del$ flag is set, then each of the traversed elements is deleted from $S$.
$\mathit{param}$ plays the same role as in \Put.
\remove{\item {\em \Transform(TraverseObject $\mathit{S}$, function $f$): returns TraverseObject}, 
which  takes as parameter a traverse object $S$ and transforms it to another traverse object $S'$
storing sets of elements of $S$. }
\end{itemize}
$S$ satisfies the {\em traversing property}: 
\yy{Each instance of \Traverse\ in every (sequential) execution of the object} 
applies $f$ at least once on all distinct elements added in $S$
and not yet been deleted by the invocation of \Traverse. 
\end{definition}

\remove{
\rb{
Definition~\ref{def:traverse} aims to provide the sequential specification of the traverse object. The
traversing property refers to every instance of \Traverse\ in every sequential execution of the object.
Every concurrent execution of the object should be linearizable, and the property must be satisfied for
the sequential execution defined by its linearization.
}
}

\begin{algorithm}[t]
\setcounter{AlgoLine}{0}
\removelatexerror
\footnotesize
\vspace*{2mm}

	\begin{flushleft}	
	\com Shared objects:\;	
\nl	TraverseObject \BC, initially containing all raw data series\;
\nl	TraverseObjects \TP, \PS, \RS, initially empty\; 
\nl	int $\mathit{BSF}$\; 
	\end{flushleft}

	\vspace*{1mm}
	\com Code for thread $t_i$, $i \in \{0, \ldots, n-1\}$:\;		
	\begin{minipage}[t]{0.45	\textwidth}
	\myproc{{{\small \QueryAnswering}(QuerySeriesSet $\mathit{SQ}$): returns int}}{
\nl	    \BC.\Traverse(\&BufferCreation(), $\mathit{BCParam}$, \False)\;
\nl	    \TP.\Traverse(\&TreePopulation(), $\mathit{TPParam}$, \False)\;
\nl		\PS.\Traverse(\&Prunning(), $\mathit{PSParam}$, \False)\;
\nl		\RS.\Traverse(\&Refinement(), $\mathit{RSParam}$, \True)\;
\nl		\Return $\mathit{BSF}$\;
	}

	\vspace*{1mm}
	\myproc{{{\small \BufferCreation}(DataSeries $\mathit{ds}$)}}{


\nl		iSAXSummary $\mathit{iSAX}$ := Calculate the iSAX summary for $\mathit{ds}$\;
\nl		Index $\mathit{bind}$ := index to appropriate buffer based on $\mathit{iSAX}$\;
\nl		\TP.\Put($\langle \mathit{iSAX}$, index of $\mathit{ds}$ $\rangle$, $\mathit{bind}$)\;
	}
	\vspace*{1mm}
	\myproc{{{\small \TreePopulation}(Summary $\mathit{iSAX}$, Index $\mathit{ind}$, 
	Index $\mathit{bind}$, Boolean $\mathit{flag}$)}}{
\nl		\PS.\Put($\langle \mathit{iSAX}$, $\mathit{ind} \rangle$, $\mathit{bind}$, $\mathit{flag}$)\;
	}
	\vspace*{1mm}
	\myproc{{{\small \Prunning}(DataSeries $Q$, DataSeriesSet $\mathit{E}$, 
	Boolean $\mathit{flag}$): returns boolean}}{

\nl		iSAXSummary $\mathit{iSAX}$ := Calculate the iSAX summary for $\mathit{E}$\;
\nl     int $\mathit{lbDist}$ := lower bound distance between $\mathit{iSAX}$ and $Q$\;
\nl	    \uIf{$\mathit{lbDist} < \mathit{BSF}$}{
\nl	       \RS.\Put($\langle \mathit{E, iSAX} \rangle$, $\mathit{flag}$)\; 
\nl		   \Return TRUE\;
		}
\nl		\Return FALSE\;
	}

	\vspace*{1mm}
	\myproc{{{\small \Refinement}(DataSeries $Q$, DataSeriesSet $E$, Summary $\mathit{iSAX}$, 
	 Function *\UpdateBSF): returns Boolean}}{
\nl		int $\mathit{lbDist}$, $\mathit{rDist}$\;


\nl	        $\mathit{lbDist}$ := lower bound distance between $\mathit{iSAX}$ and $Q$\;
\nl	        \uIf{$\mathit{lbDist} < \mathit{BSF}$}{
\nl			\For{\Each pair $\langle \mathit{iSAX_{ds}, ind_{ds}} \rangle$ in $E$} {
\nl				$\mathit{lbDist}$ := lower bound distance between $\mathit{iSAX_{ds}}$ and $Q$\;
\nl				\uIf {$\mathit{lbDist} < \mathit{BSF}$} { 
\nl					$\mathit{rDist}$ := real distance between $\mathit{ds}$ and $Q$\;
\nl					\uIf {$\mathit{rDist} < \mathit{BSF}$} { 
\nl						*\UpdateBSF($\mathit{BSF},\mathit{rDist}$) \com user-provided routine\; 
					}
				}
			}
\nl			return \True
		}
\nl		\lElse{ return \False
		}
	}
	\end{minipage}	
\caption{Implementation of an iSAX-based index using the traverse objects \BC, \TP, \PS, \RS.}
\label{alg:iSAXTraverse}
\end{algorithm}


We use four instances of a traverse object to implement the four stages of an iSAX-based index.
We call \BC, \TP, \PS, and \RS, the traverse objects that implement
the buffer creation, tree population, prunning, and refinement stages, respectively.
The buffers creation phase uses an array \RawData\ to store
the raw data series, 
thus, the elements of $\mathit{BC}$ are stored in \RawData.
The tree population phase uses a set of arrays ({\em summarization buffers}) 
where the pairs  of iSAX summaries 
and pointers to data series are initially stored. 
\TP\ stores these pairs. The prunning stage 
employs a leaf-oriented tree to store these pairs. 
Thus, \PS\ organizes the pairs into as many sets as the leaf nodes of the tree. 
Each set contains the pairs stored in each leaf. Finally, the refinement stage
uses priority queues to store those tree leaves containing candidate series. 

Answering a query 
is now comprised of a sequence of four invocations of \Traverse\ on the different traverse objects.
Algorithm~\ref{alg:iSAXTraverse} provides pseudocode for the implementation of an iSAX-based index using traverse objects. 
%
%
The four stages of an iSAX-based index
do not overlap with one another. This is usually ensured with the use of 
synchronization barriers. In the scheme of Algorithm~\ref{alg:iSAXTraverse},
the barriers, if needed, (as well as multithreading processing) 
should be incorporated in the implementation of \Put\ and \Traverse. 
Thus, an iSAX-based index satisfies the following property.

\begin{definition}[Non-Overlapping Property]
\label{def:tr-prop}
\yy{In every (concurrent) execution of the index and every traverse object $S$ accessed in the execution,} 
each instance of \Traverse\ on $S$  
is performed only after the execution of all instances of \Put\ that add distinct
elements in $S$ has been completed. 
\end{definition}

\remove{\rb{
Definition~\ref{def:tr-prop} refers to concurrent executions of an index, stating that the invocation of \Traverse\ on
every traverse object $S$ in them must appear after the response of any PUT invoked on $S$.
}
}

Assume that the non-overlapping property holds for \BC, \TP, \PS, and \RS\
and that \RawData\ initially stores all raw data series. 
The traversing property implies that the \BufferCreation\ function 
is invoked
at least once for each data series $\mathit{ds}$ in \RawData, 
so at least one appropriate pair is added
for it in \TP, i.e., 
the summarization buffers 
are populated appropriately. 
By the non-overlapping 
and the traversing  
properties,
\TreePopulation\ is invoked for all these pairs. 
Since \TreePopulation\ invokes \Put\ on \PS, it follows that 
at least one pair for each of the data series of the collection 
is added in \PS\ (i.e. in the tree index). 
By the traversing property, 
all elements of \PS\ 
are traversed and \Prunning\ is called
on them. Thus, all tree leaves that cannot be pruned are added in \RS. Note that 
\Traverse\ on \RS\ is invoked with the $\mathit{del}$ flag being  \True. 
This allows to use (one or more) priority queues for implementing \RS, and to employ \DeleteMin\ 
to delete each traversed element during \Traverse. \Refinement\ 
will be applied on every traversed element of \RS. Therefore, those leaves
that cannot be pruned will be further processed by calculating real distances
and for the data series they store, and by updating $\mathit{BSF}$ whenever needed. 
%
Implementations for \Put\ and \Traverse\ for $BC$, $TP$, $PS$, and $RS$ in \Fresh\ 
are presented in Section~\ref{sec:fresh}. 
\remove{Section~\ref{sec:ds} provides implementations
of data structures with specific functionality and other mechanisms that 
comprise major components of the implementations in Section~\ref{sec:fresh}. 
We believe that some of these data structures
can be of independent interest (e.g., be useful for building DS indexes that are
not iSAX-based or in other big-data applications).

\here{Explain that PS and RS could be called repeatedly to answer a batch of queries.
See the foreach in code of query answering in Algorithm 1 which is commented out.}
}

\section{Locality-Aware Lock-Freedom}
\label{sec:llf}

\begin{figure*}[tb]
	\centering
	\includegraphics[width=0.9\textwidth]{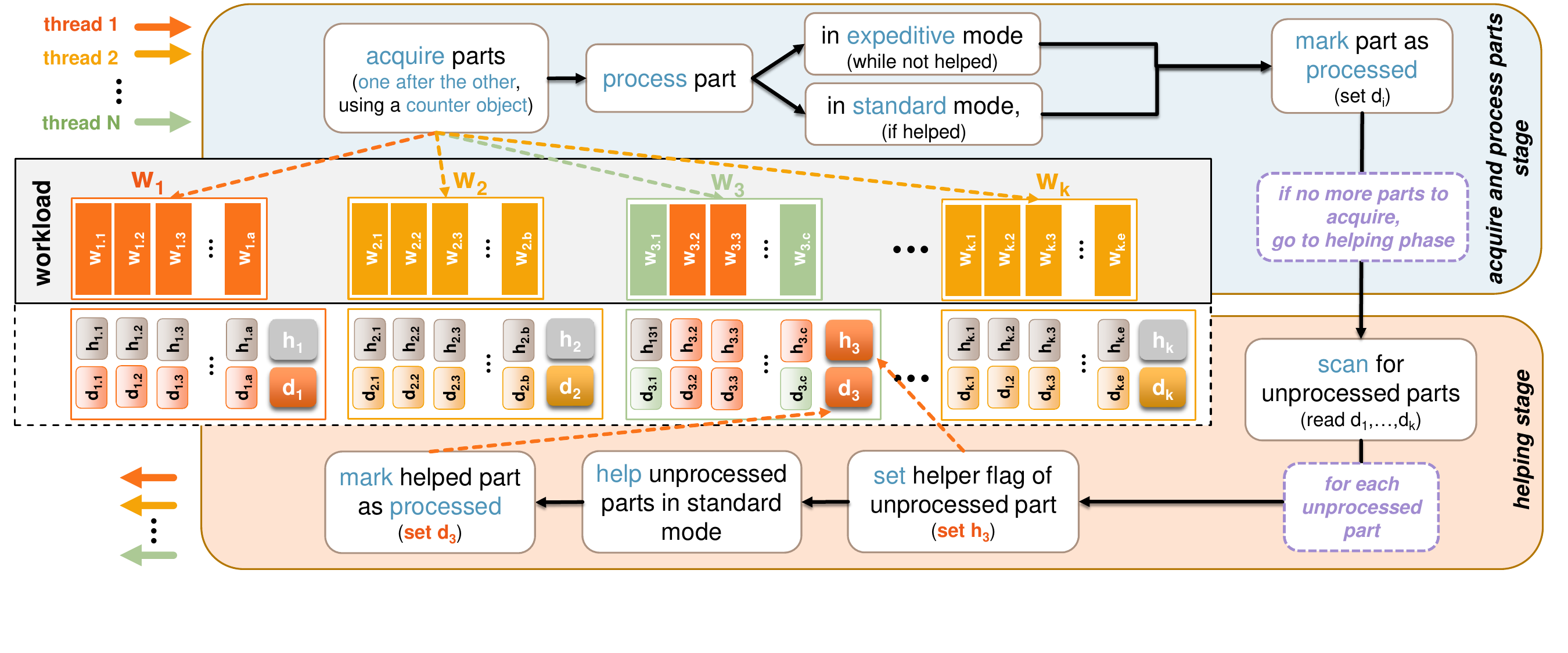}	
	\vspace*{-0.8cm}
	\caption{Refresh flowchart.}
	\vspace*{-0.2cm}
	\label{fig:refresh:flowchart}
\end{figure*}

\remove{We present a generic methodology, called \Refresh, which can 
be used to transform {\em blocking} implementations 
of iSAX-based indexes into lock-free ones that are as highly efficient
as their blocking analogs.

We first introduce the notion of {\em locality-awareness}. }
{\em Locality-awareness} 
aims at capturing several design principles (Definition~\ref{def:principles})
for data series indexes which are crucial for 
achieving good performance.
A locality aware implementation 
respects these principles. 

\begin{definition}
\label{def:principles}
Principles for {\em locality-aware} processing:
\begin{enumerate}
\item
{\bf Data Locality.} Separate the data into {\em disjoint sets} and have a distinct thread
processing the data of each set. 
This results in 
reduced communication cost (cache misses and branch misprediction) among the threads.
\item
{\bf High Parallelism \& Low Synchronization Cost.} 
Threads should work in parallel and independently from each other. 
Whenever synchronization cannot be avoided, design 
mechanisms to minimize its cost.
\item
{\bf Load Balancing.} Share the workload equally to the different threads, thus
avoiding load imbalances between threads and having 
all threads busy at each point in time. 
\end{enumerate}
\end{definition}

Enuring locality awareness results in good performance and is thus 
a desirable property for big data processing. 
In existing iSAX-based indexes, a thread operates on
chunks of \RawData\ and 
processes disjoint sets of summarization
buffers and subtrees of the index tree. 
Also, an iSAX-based index employs several priority queues
to store leaf nodes containing candidate series.
Thus, 
iSAX-based indexes are {\em locality-aware}.
%

To describe \Refresh\ in more detail,
consider a {\em blocking} locality-aware
implementation $\mathcal{A}$, which 
splits its workload 
into disjoint parts and assigns them  to 
threads for processing. 
%
%
%
\Refresh\ (Algorithm~\ref{alg:recipe}) transforms $\mathcal{A}$ into a {\em lock-free} locality-aware implementation 
$\mathcal{B}$ that achieves high parallelism. 

Let $W$ be the workload that $\mathcal{A}$ processes 
and let $w_1, \ldots, w_k$ be the parts it is separated to ensure locality awareness. 
\Refresh\ applies
the following steps \yy{(depicted in Figure~\ref{fig:refresh:flowchart})}:

\begin{algorithm}[t]
	\setcounter{AlgoLine}{0}
		\removelatexerror
		\footnotesize
		\begin{flushleft}	
			\com Shared variables:\;		
\nl			workload part $\mathit{W}$ := $\mathit{[w_1, w_2, \ldots, w_k}]$ \label{alg:recipe:r}\;
\nl			\Boolean\ $\mathit{F}$ := $\mathit{[d_1, d_2, \ldots, d_k]}$, initially $\mathit{d_i} = \False$, $\mathit{1 \leq i \leq k}$ \label{alg:recipe:c}\;		
\nl			\Boolean\ $\mathit{H}$ := $\mathit{[h_1, h_2, \ldots, h_k]}$, initially $\mathit{h_i} = \False$, $\mathit{1 \leq i \leq k}$ \label{alg:recipe:h}\;
		\end{flushleft}

		\vspace*{1mm}

		\com Code for each thread:\;		
		\myproc{{{\small \Refresh}()}}{
\nl	   		\tcp{acquire and process parts of $\mathit{W}$}
			\While{$\mathit{W}$ has available parts}{	\label{alg:recipe:process:start}
				$\mathit{w_i} := $ {\bf acquire} an available part of $\mathit{W}$ \;
				{\bf mark} $\mathit{w_i}$ as acquired \;
				\uIf{$\mathit{h_i = \False}$}{ \label{alg:recipe:process:if}
					{\bf process} $\mathit{w_i}$ in {\bf expeditive} mode, while checking that $\mathit{h_i}$ remains $\mathit{\False}$; in case $\mathit{h_i = \True}$, {\bf switch} to {\bf standard} mode \label{alg:recipe:process:expeditive}
				} 
	   			\lElse{{\bf process} $\mathit{w_i}$ in {\bf standard} mode}	\label{alg:recipe:process:standard}
	   			$\mathit{d_i}$ := $\mathit{\True}$ \label{alg:recipe:c:true}
	   		}
	   		\tcp{scan flags for unfinished parts of $\mathit{W}$ and help}
	   		\For{\Each $\mathit{d_i \in D}$ with $d_i = \False$} { \label{alg:recipe:scan:for} 
	   			\Backoff()	\tcp*{avoid helping, if possible}				\label{alg:recipe:help:backoff}
	   			\uIf{$\mathit{d_i = \False}$}{								\label{alg:recipe:help:if}
	   				$\mathit{h_i := \True}$ \;								\label{alg:recipe:h:true}
		   			{\bf process} $\mathit{w_i}$ in {\bf standard} mode, while periodically checking that $\mathit{d_i}$ remains $\mathit{\False}$; in case $\mathit{d_i} = \mathit{\True}$, {\bf stop} processing $\mathit{w_i}$ \label{alg:recipe:help:process}\;
					$\mathit{d_i := \True}$											\label{alg:recipe:help:c:true}
		   		}
	   		}
		}

		\vspace*{1mm}

	\caption{\Refresh - A general approach for transforming a blocking data structure $\mathit{D}$ of a big-data application $\mathcal{A}$ 
into a lock-free one.}
	\label{alg:recipe}
\end{algorithm}

\noindent
{\bf (1)} It attaches a {\em flag} $d_i$, $1 \leq i \leq k$, (initially \False)
with each $w_i$ to identify whether $w_i$'s processing 
is done. 
As soon as a thread
finishes processing $w_i$, it sets $d_i$ to \True\ (line~\ref{alg:recipe:c:true}).

\noindent
{\bf (2)} Threads in $\mathcal{B}$ execute the same algorithm as 
in $\mathcal{A}$ to acquire parts of $W$ to process,
until all parts have been acquired (lines~\ref{alg:recipe:process:start}-\ref{alg:recipe:c:true}). 
The thread that acquires a workload is its {\em owner}.

\noindent
{\bf (3)} 
To achieve lock-freedom,
every thread $t$, then, {\em scans} all the flags 
to find those parts that are still unfinished
(line~\ref{alg:recipe:scan:for}).

\noindent
{\bf (4)} Thread $t$ {\em helps} by processing, one after the other, each part found unfinished
during scan. 
For each part $w_i$ 
that $t$ helps, it periodically checks $d_i$
to see whether other threads completed the processing of $w_i$. If this is so,
$t$ stops helping $w_i$ (line~\ref{alg:recipe:help:process}). 
A thread that completes the processing of $w_i$, changes $d_i$ to true (line~\ref{alg:recipe:help:c:true}).

\noindent
{\bf (5)} Due to helping, 
every data structure $D$, employed in $\mathcal{A}$, may be 
concurrently accessed by many threads. 
Thus, $\mathcal{B}$ should provide an efficient 
lock-free implementation for all data structures of $\mathcal{A}$. 

In locality-aware implementations, 
threads are expected to work on their own parts of the data 
most of the time ({\em contention-free phase}), and they may help other threads only for a small period of time at the end of their execution
({\em concurrent phase}).
In the contention-free phase, \Refresh\ 
avoids synchronization overheads incurred to ensure lock-freedom. 
%
Specifically, it employs two implementations for each data strucutre $D$ of  $\mathcal{A}$, 
one with low synchronization cost that does not support helping ({\em expeditive mode}),
and another that supports helping and has higher synchronization overhead ({\em standard mode}). 
To enable threads operate on the appropriate mode,  a {\em helping-indicator flag} $h_i$ (initially \False) 
is attached with each $w_i$. 
A thread $t$ starts by processing its assigned workload 
on expeditive mode (lines~\ref{alg:recipe:h} and \ref{alg:recipe:process:if}-\ref{alg:recipe:process:expeditive}).
Before $t$ starts helping some part $w_i$, it sets $h_i$ to \True\
(line~\ref{alg:recipe:h:true}),
to alert $w_i$'s owner thread to start running on standard mode (line~\ref{alg:recipe:process:expeditive}).

To avoid helping whenever it is not absolutely necessary, 
\Refresh\ provides an optional {\em backoff scheme} that is used
by every thread $t$ (line~\ref{alg:recipe:help:backoff}) 
before it attempts to help other threads (line~\ref{alg:recipe:help:if}-\ref{alg:recipe:help:process}).
A small delay before switching to standard mode, 
often positively affects performance. The delay is usually an estimate 
of the actual time a thread requires to finish its current workload, 
calculated at run time (see Section~\ref{sec:bc} for more details).
%
To minimize the work performed by a helper, \Refresh\ could
be applied  {\em recursively} by splitting each part $w_i$ 
to subparts. This way, a helper helps only the remaining unfinished subparts
of $w_i$.

Lock-freedom is ensured due to the {\em helping code} (lines~\ref{alg:recipe:scan:for}-\ref{alg:recipe:help:c:true}).
In \Refresh, only after a thread $t$ processes a workload $w_i$, 
it sets $h_i$ to \True\, 
and $t$ performs the helping code after finishing with their assigned workloads.  
Thus, when $t$ completes its helping code
the processing of all parts of the workload has been completed. 
This means that $t$ may continue directly to the
execution of the next stage, without waiting for the other threads to complete the execution of the current stage. 
Therefore, this scheme renders the use of barriers useless, as needed to achieve lock-freedom. 
\remove{\begin{theorem}
\label{thm:refresh}
\Refresh\ is a general scheme for processing a locality-aware workload in a lock-free way, without sacrificing locality-awareness. 
\end{theorem}
}

Summarizing, \Refresh\ is a general scheme for processing a locality-aware workload in a lock-free way, without sacrificing locality-awareness. 

\section{FreSh}
\label{sec:fresh}

We follow the data processing flow, described in Section~\ref{sec:traverse},
employ $\mathit{BC}$, $\mathit{TP}$, $\mathit{PS}$, and $\mathit{RS}$ (from Section~\ref{sec:traverse}),
and repeatedly apply \Refresh\ (from Section~\ref{sec:llf})
to come up with \Fresh. 
%
\remove{\Fresh\ 
employs the four traverse objects, 
$\mathit{BC}$, $\mathit{TP}$, $\mathit{PS}$, and $\mathit{RS}$ (see Section~\ref{sec:traverse}), 
to implement the buffer creation, the tree population, the pruning and the
refinement stages, respectively. 
}

\remove{\BC\ is implemented using a single buffer, called
\RawData, \TP\ is implemented using a number of $k$ {\em summarization buffers}, 
\PS\ is implemented using a leaf-oriented tree, and \RS\ is implemented using
a set of priority queues, each implemented with a sorted array. 
Its code follows the general scheme provided in Algorithm~\ref{alg:iSAXTraverse}. 
We now present the details of the implementation of these traverse objects. 
}

\subsection{Buffers Creation and Tree Population} 
\label{sec:bc}

\BC\ is implemented using a single buffer, called
\RawData. 
In \BC, \Put\ 
is never used, as we assume that the data are initially in \RawData.
To implement \Traverse, we employ \Refresh. 
We split \RawData\ into $k$ equally-sized {\em chunks} of consecutive 
elements 
to get 
$k$ workloads. 
Threads use a counter object to get assigned chunks to process.
To reduce the cost of helping, \Fresh\ calls \Refresh\ recursively.
Specifically, it splits each chunk into smaller parts,
called {\em groups}, and employ \Refresh\ a second time for processing the groups of a chunk.
In more detail, 
\Fresh\ maintains an additional counter object 
for each chunk of $RawData$. 
Each thread $t$ that acquires or helps a chunk, uses the counter object of the chunk to acquire {\em groups} in 
the chunk to process. 
\Fresh\ also applies a third level of \Refresh\ recursion,
where each workload is comprised of the processing of just a single element
of a group. 

Pseudocode for \BC.\Put\ and \BC.\Traverse\ is provided 
in Algorithm~\ref{alg:bc}. \RawData\ is comprised of $k$ chunks,
each containing $m$ groups; moreover, each group contains $r$ elements (line~\ref{alg:bc:r}). 
\Fresh\ uses three
sets of done flags, $\mathit{DChunks}$, $\mathit{DGroups}$,
and $\mathit{DElements}$ (line~\ref{alg:bc:c}), storing 
one done flag for each chunk,
for each group, and 
for each element, respectively.
Similarly, \Fresh\ employs three sets of counter objects,
$\mathit{Chunks}$, $\mathit{Groups}$,
and $\mathit{Elements}$ (line~\ref{alg:bc:e})), to count the chunks, groups and elements, assigned to threads for processing. 
\Fresh\ also employs two sets of {\em helping} flags (line~\ref{alg:bc:h}), 
$\mathit{HChunks}$  (for helping chunks) and $\mathit{HGroups}$ (for helping groups). 
%
In an invocation of
\Traverse(\&\BufferCreation, \RawData, $\mathit{Dchunks}$, $\mathit{DGroups}$, $\mathit{DElements}$, $\mathit{HChunks}$, $\mathit{HGroups}$, 
\False, $\mathit{Chunks}$, $\mathit{Groups}$, $\mathit{Elements}$, $1$),
$h$ is equal to \False. By the way a counter object works, it follows that 
no expeditive mode is ever executed at the first level of the recursion.
Note that at this level, the roles of $D_1$ and $H_1$ are played by the one-dimensional arrays $DChunks$ and $HChunks$, 
respectively. Moreover, $DGroups$ and $DElements$ play the role of $D_2$ and $D_3$,
respectively, and $HGroups$ plays the role of $H_2$.
Each chunk is processed by recursively calling \Traverse\ ({\em level-2 recursion}) 
on line~\ref{alg:bc:recur} (with $\mathit{rlevel}$ being equal to $2$). 
The goal of a level-2 invocation of \Traverse\
is to process an entire chunk by splitting it into groups and calling
\Traverse\ once more ({\em level-3 recursion}) to process the elements of each group 
(recursive call of line~\ref{alg:bc:recur} with $\mathit{rlevel}$ being equal to $3$). 
Note that in a level-2 invocation corresponding to some chunk $i$, 
\RawData\ is the two-dimensional array containing the elements of the groups of chunk $i$.
Moreover, 
the role of $D_1$ is now played by the one-dimensional array $DGroups[i]$, 
and the role of $D_2$ by the two-dimensional array $DElements[i]$,
whereas $D_3$ is no longer needed and is $\mathit{NULL}$.
The role of $H_1$ is now played by the one-dimensional array $HGroups[i]$. 
Helping (lines~\ref{alg:bc:scan:for}-\ref{alg:bc:help:c:true}) follows the general pattern
described in Algorithm~\ref{alg:recipe}. 


\begin{algorithm}[t]
	\setcounter{AlgoLine}{0}
		\removelatexerror
		\footnotesize
		\begin{flushleft}	

			\com Shared variables:\;		
\nl			Set $\mathit{RawData}[1..k][1..m][1..r]$, initially containing all data series \label{alg:bc:r} \;
		\vspace*{1mm}
\nl			\Boolean\ $\mathit{DChunks[1..k]}$, $\mathit{DGroups[1..k][1..m]}$,  \\ \hspace*{.8cm}$\mathit{DElements}[1..k][1..m][1..r]$,  initially all \False \label{alg:bc:c}\;		
		\vspace*{1mm}
\nl			\Boolean\ $\mathit{HChunks}[1..k]$, $\mathit{HGroups}[1..k][1..m]$, initially all \False \label{alg:bc:h}\;

		\vspace*{1mm}
\nl		CounterObject $\mathit{Chunks}$, $\mathit{Groups[1..k]}$, $\mathit{Elements[1..k][1..m]}$ \label{alg:bc:e}\;
\nl		int $\mathit{Size[1..3] = \{k,m,r\}}$\;
		\vspace*{1mm}
	\end{flushleft}
%

		\myproc{{{\small \Traverse}(Function *BufferCreation, DataSeries RawData[], Boolen $D_1[]$, Boolean $D_2[]$, Boolean $D_3[]$, Boolean $H_1[]$, Boolean $H_2[]$, Boolean $h$, CounterObject $Cnt_1$, CounterObject $Cnt_2[]$, CounterObject $Cnt_3[]$, int $\mathit{rlevel}$)}}{
\nl			int $\mathit{i}$\;
	   		\tcp{acquire and process parts of $\mathit{W}$}
			\While{$\True$}{
				$\mathit{\langle i, * \rangle := Cnt_1.\NextIndex(\&h)}$\;
				\lIf{$\mathit{i > Size[rlevel]}$}{\Break}
				{\bf mark} $\mathit{\RawData[i]}$ as acquired \;
				\lIf{$\mathit{rlevel < 3}$}{				
		   			\Traverse($\mathit{BufferCreation, \RawData[i], D_2[i], D_3[i]}$,
					\hspace*{.2cm} $\mathit{D_3[i], H_2[i], NULL, H_1[i], Cnt_2[i], Cnt_3[i], Cnt_3[i])}$
					\hspace*{.2cm} $\mathit{ rlevel+1)}$
					\label{alg:bc:recur} 
				}
				\lElse{ \label{alg:sb:em:else}
					*\BufferCreation($\RawData[i]$)\;	
				}
	   			$\mathit{D_1[i]}$ := $\mathit{\True}$ \label{alg:bc:c:true}
	   		}
	   		\tcp{scan flags for unprocessed parts of $\mathit{W}$ and help}
	   		\For{\Each $j$ such that $\mathit{D_1[j]}$ is equal to \False} { \label{alg:bc:scan:for} 
	   			\Backoff()	\tcp*{avoid helping, if possible}				\label{alg:bc:help:backoff}
	   			\uIf{$\mathit{D_1[j]] = \False}$}{						\label{alg:bc:help:if}
	   				$\mathit{H_1[j] := \True}$ \;						\label{alg:bc:h:true}
					\lIf{$\mathit{rlevel < 3}$}{
			   			\Traverse($\mathit{BufferCreation,\RawData[j], D_2[j], D_3[j]}$,
						\hspace*{.1cm} $\mathit{ D_3[j],H_2[j], NULL, H_1[j], Cnt_2[j], Cnt_3[j]}$,
						\hspace*{.1cm} $\mathit{ Cnt_3[j]), rlevel+1)}$\label{alg:bc:help:process}
					} 
					\lElse{ \label{alg:sb:em:else}
						*\BufferCreation($\RawData[j]$)	
					} 
				$\mathit{D_1[j] := \True}$\label{alg:bc:help:c:true}
		   	}
			}			
	   	}
		\vspace*{1mm}

	\caption{Pseudocode for \Traverse\ of \BC\ in \Fresh. Code for thread $t$.}
	\label{alg:bc}
\end{algorithm}

The backoff time in \Fresh\
depends on the average execution time required by each thread to process a group.
Each thread $\mathit{t}$ counts the average time $T_{avg}$ it has spent to process 
all the parts it acquired, and whenever it encounters a group to help, it 
sets the backoff time to be proportional to $T_{avg}$ and performs helping
only after backoff, if it is still needed. 

\Fresh\ implements \TP\ using a set of $2^w$ summarization buffers
($w$ is the number of segments of an iSAX summary),
one for each bit sequence of $w$ bits.
To decide to which summarization buffer to store a pair,
\Fresh\ 
examines the bit sequence consisting of the first bit of each 
of the $w$ segments of the pair's iSAX summary, 
and places the pair into the corresponding summarization buffer. 
Each of the summarization buffers is split into $N$ parts, one for each of the $N$ threads in the system.
Each thread uses its own part in each buffer to store the elements it inserts.

To implement \TP.\Traverse, we split the elements of \TP\ into $2^w$ workloads 
and apply \Refresh. 
Each summarization buffer could be further split into chunks and groups, and \Refresh\ could be called recursively.
Pseudocode for \Traverse\ of \TP\ closely follows that for \BC.
%
\BC\ and \TP\ are lock-free implementations of a traverse object. 

\subsection{Prunning and Refinement}

In \Fresh, \PS\ is implemented as a 
forest of $2^w$ leaf-oriented trees, 
one for each of the summarization buffers. 
The trees of the forest are the root subtrees of a standard iSAX-based tree.
%
To implement \PS.\Traverse, \Fresh\ uses \Refresh\ to process 
the different subtrees of the index tree. Specifically, 
each thread $t$ access a counter to get assigned a subtree $T$ to process. 
To process the nodes of $T$, \Refresh\ is applied recursively. 
A thread $t$ that is assigned node $i$ of $T$,
first searches for the $i$-th node, according to inorder,
and then processes it by invoking the \Prunning\ function 
of Algorithm~\ref{alg:iSAXTraverse}. 
%
To find the $i$-th node of $T$
in  an efficient way, 
for each node $nd$ of $T$, \Fresh\ maintains a counter
$\mathit{cnt_{nd}}$ that counts the number of nodes in the left subtree
of $\mathit{nd}$. \Fresh\ 
uses these counters to find the $i$-th node of $T$ by simply traversing a path in $T$.
The total number of nodes in T is calcualted by
simply traversing the righmost path of $T$ and summing up the counters stored in the 
traversed nodes. 

\subsubsection{Insert in Leaf-Oriented Tree}
\label{sec:trees}
Each node of the tree stores a key and
the pointers 
to its left and right children. 
A leaf node stores additionally an array $D$, where the leaf's data are stored. 
We assume that each data item is a pair containing a key and the associated information.
A node may have its own key. For instance, in iSAX-based indexes, 
this key is the node's iSAX summary. 
The proposed implementation allows multiple 
insert operations to concurrently update array $D$ of a leaf. 
This results in enhanced parallelism and performance. 
To achieve this, each leaf $\ell$ contains a counter, called $\mathit{Elements}$.
Each thread $\mathit{t}$ that tries to insert
data in $\ell$, uses $\mathit{Elements}$ to acquire a position $\mathit{pos}$ in the array $D$ of $\ell$.
If $D$ is not full, 
$\mathit{t}$ stores the new element in $D[\mathit{pos}]$. 
Otherwise, 
$\mathit{t}$ attempts to split the leaf. 

During spliting, $D$ may contain empty positions, since some
threads may have acquired positions in $D$
but have not yet stored their elements there. 
To avoid situations of missing elements, 
each leaf contains an $\mathit{Announce}$ array 
with one position for each thread. A thread announces its operation
in $\mathit{Announce}$ before it attempts to acquire a position in $D$.
During spliting, a thread distributes to the new leaves it creates not only the
elements found in $D$ but also those in $\mathit{Announce}$.

Our approach is a {\em linearizable, lock-free} implementation of a leaf-oriented
tree with fat leaves (supporting only insert).

\remove{
\begin{lemma}
\label{lem:tree}
Algor.~\ref{alg:tree} is a {\em linearizable, lock-free} implementation of a leaf-oriented
tree with fat leaves, supporting only insert operations.
\end{lemma}
}

\subsection{Refinement}

To implement \RS, \Fresh\ uses a set of priorities queues
each implemented using an array. 
A thread inserts elements in all arrays in a round-robin fashion. 
This technique results in almost equally-sized arrays, which is crucial
for achieving load-balancing. 


\remove{
During query answering, an application may use the same index tree
to answer more than one query. In that case, the done flags of the nodes and other
variables need to be reset each time a new query starts. This may require syncrhonization. 
To avoid this cose, \Fresh\ implements the {\em done} flag as a counter (rather than as a boolean). 
This counter describes the number of queries for which the node has been processed.

\noindent
{\bf Refinement phase.}
}
To implement \RS.\Traverse, 
\Fresh\ first 
comes up with sorted versions of the arrays, shared to all threads.
Then, it uses \Refresh\ to assign 
sorted arrays to threads for processing. 
To process the elements of a sorted array $SA$, \Refresh\ is applied recursively. 
Processing of an array element is performed by invoking the \Refinement\ function (Algorithm~\ref{alg:iSAXTraverse}). 
Helping is done at the level of 1) each individual priority queue and 
2) the set of priority queues, in a way similar to that in \PS. 
\RS\ is a linearizable lock-free implementation of a traverse object. 

To update BSF, \Fresh\ repeatedly reads the current value $y$ of BSF, and attempts to atomically 
change it from $v$ to the new value $v'$, 
using \CAS, until it either succeeds or some value 
smaller than or equal to $v'$ is written in BSF.


\remove{
\begin{lemma}
\label{lem:rs}
{\bf (1)} \RS\ is a linearizable lock-free implementation of a traverse object that supports \Put\ and
\Traverse($\mathit{*, *, 0}$). 
{\bf (2)} For every thread $t$, when an invocation of \Traverse(\&$\mathit{\Prunning, *, \True}$)\ by $t$ on \RS\ completes, 
for every leaf $\ell$ in $\RS$, $\ell$ either has been processed or it has been pruned.
\end{lemma}
}


\begin{theorem}
\label{thm:qa}
\Fresh\ solves the 1-NN problem and provides a lock-free implementation of \QueryAnswering\  (Alg.~\ref{alg:iSAXTraverse}). 
\end{theorem}

\section{Experimental Evaluation}
\label{section:evaluation}

\begin{figure*}[t]
	\centering
	\begin{subfigure}[c]{0.24\textwidth}
		\includegraphics[width=1\textwidth]{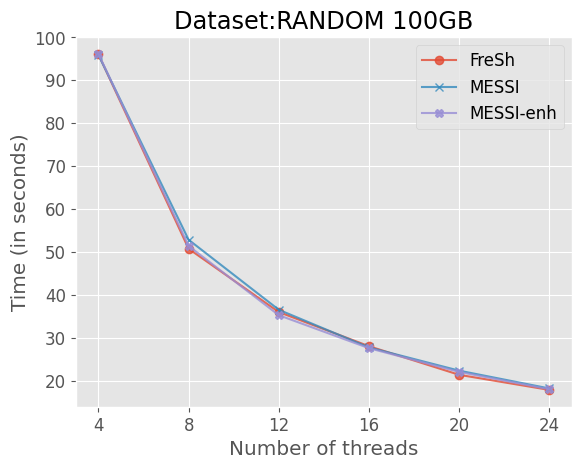}
		\caption{Total}
		\label{fig:eval:fresh-messi-threads:random:total-from-4}
	\end{subfigure}		
	\begin{subfigure}[c]{0.24\textwidth}	
		\includegraphics[width=1\textwidth]{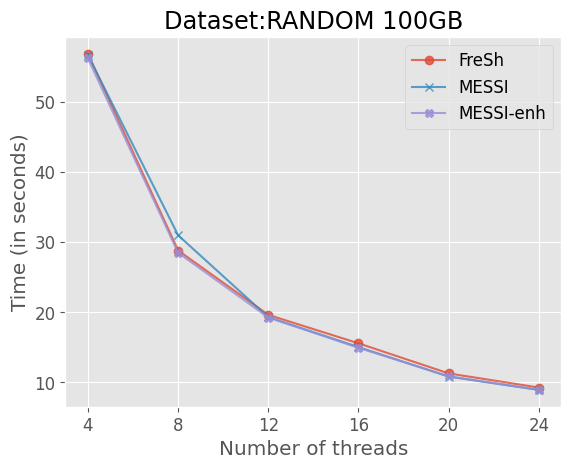}
		\caption{Summarization}
		\label{fig:eval:fresh-messi-threads:random:recbuf-from-4}
	\end{subfigure}		
	\begin{subfigure}[c]{0.24\textwidth}	
		\includegraphics[width=1\textwidth]{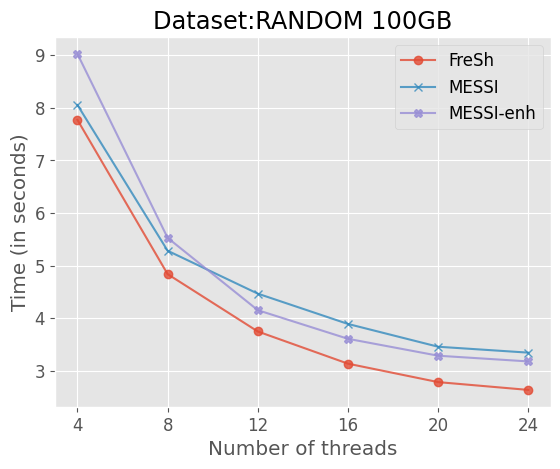}
		\caption{Tree index}
		\label{fig:eval:fresh-messi-threads:random:tree-from-4}
	\end{subfigure}		
	\begin{subfigure}[c]{0.24\textwidth}	
		\includegraphics[width=1\textwidth]{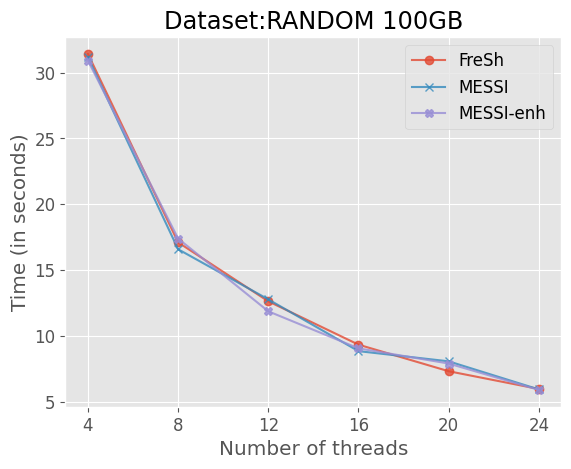}
		\caption{Query answering}
		\label{fig:eval:fresh-messi-threads:random:queries-from-4}
	\end{subfigure}		
	\vspace*{-0.2cm}
	\caption{Comparison of \Fresh\ 
	against \MESSI\ and \MESSIenh\ on 100GB Random. 
	}
	\label{fig:eval:fresh-messi-threads:seismic}
\end{figure*}

\begin{figure*}[h]
	\begin{subfigure}{0.24\textwidth}
		\includegraphics[width=1\textwidth]{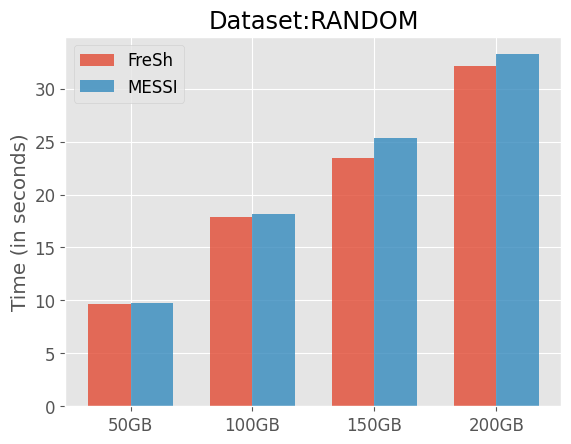}
		\caption{Total}
		\label{fig:eval:scale-dataset:random:total}
	\end{subfigure}		
	\begin{subfigure}{0.24\textwidth}
		\includegraphics[width=1\textwidth]{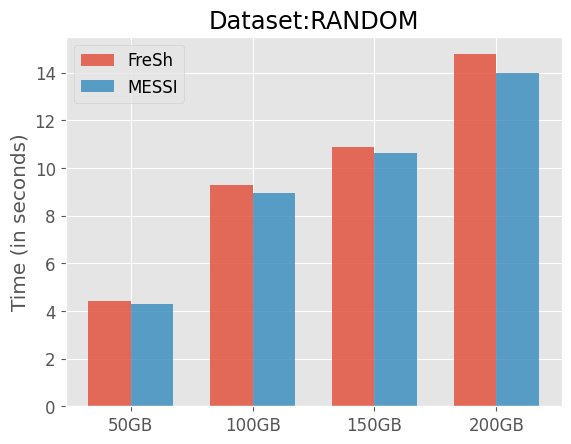}
		\caption{Summarization}
		\label{fig:eval:scale-dataset:random:summarization}
	\end{subfigure}		
	\begin{subfigure}{0.24\textwidth}
		\includegraphics[width=1\textwidth]{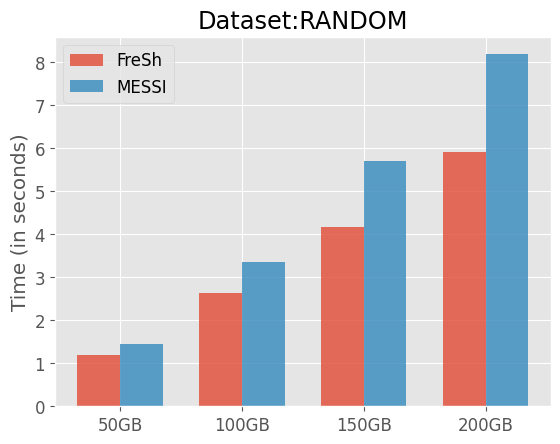}
		\caption{Tree index}
		\label{fig:eval:scale-dataset:random:tree-index}
	\end{subfigure}		
	\begin{subfigure}{0.24\textwidth}
		\includegraphics[width=1\textwidth]{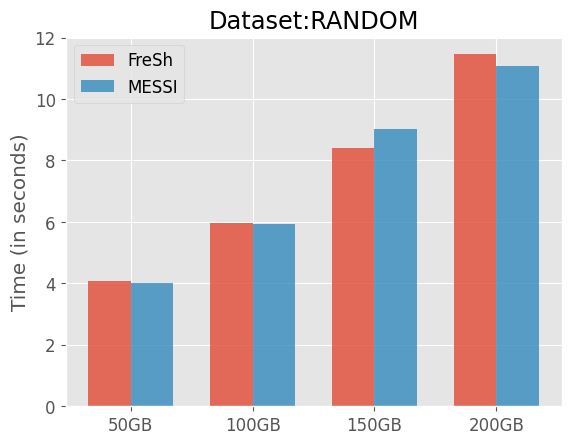}
		\caption{Query answering}
		\label{fig:eval:scale-dataset:random:query-answering}
	\end{subfigure}				
	\vspace*{-0.2cm}
	\caption{Comparison of \Fresh\ against \MESSI: (a-d) on Random, 
	for 24 threads.}
	\label{fig:eval:scale-dataset:seismic}
\end{figure*}

\begin{figure}[h]	
	\begin{subfigure}{0.24\textwidth}
		\includegraphics[width=1\textwidth]{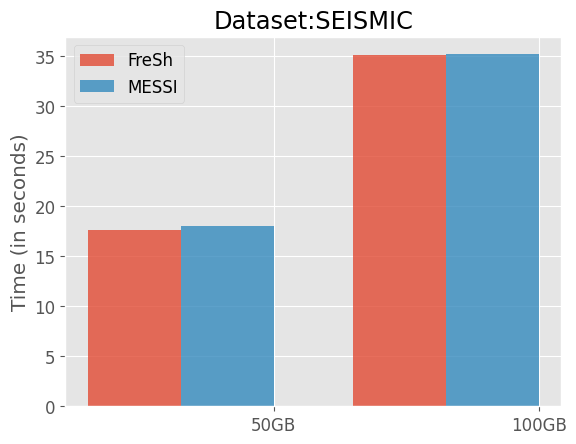}
		\caption{Total}
		\label{fig:eval:scale-dataset:astro:total}
	\end{subfigure}		
	\begin{subfigure}{0.23\textwidth}
		\includegraphics[width=1\textwidth]{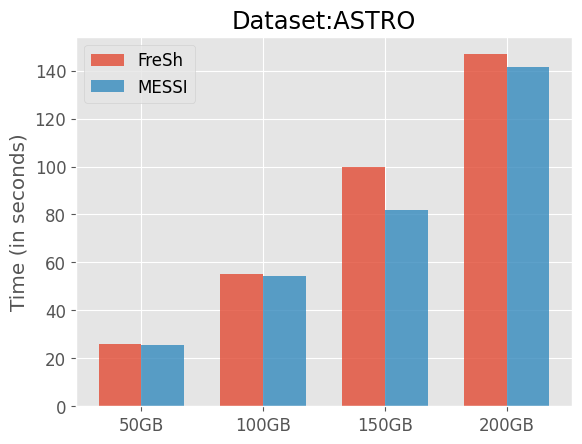}
		\caption{Total}
		\label{fig:eval:scale-dataset:seismic:total}
	\end{subfigure}		
	\vspace*{-0.2cm}
	\caption{Comparison of \Fresh\ against \MESSI: 
	(a) on Seismic,  
        and (b) on Astro, 
	for 24 threads.}
	\vspace*{-0.4cm}
	\label{fig:eval:scale-dataset:seismic}
\end{figure}

\noindent{\bf Setup.}
\remove{
\lc{We should describe nefeli here.}
We used a 40-core machine equipped with 4 Intel(R) Xeon(R) E5-4610 v3 1.7Ghz CPUs
with 10 cores each, 
and 25MB L3 cache. 
The machine runs CentOS Linux 7.9.2009
with kernel version 3.10.0-1160.45.1.el7.x86\_64 and has 256GB of RAM.
Code is written in C and compiled using the gcc compiler (version 11.2.1) 
with O2 optimizations. 
}
\y{
We used a machine equipped with 2 Intel  Xeon E5-2650 v4 2.2GHz
CPUs with 12 cores each, and 30MB L3 cache. The machine runs
Ubuntu Linux 16.04.7. LTS and has 256GB of RAM. Code is written in C and compiled using gcc v11.2.1) with O2 optimizations.}

\noindent{\bf Datasets.}
We evaluated \Fresh\ and the competing algorithms (all algorithms are in-memory) using both real and synthetic datasets.
The synthetic data series, \emph{Random}, are generated as random-walks (i.e., cumulative sums) of 
steps that follow a Gaussian distribution (0,1).
This type of data has been extensively used~\cite{conf/sigmod/Faloutsos1994,isax2plus,conf/kdd/Zoumpatianos2015,DBLP:journals/vldb/ZoumpatianosLIP18,DBLP:journals/pvldb/EchihabiZPB18,DBLP:journals/pvldb/EchihabiZPB19}, 
and models the distribution of stock market prices~\cite{conf/sigmod/Faloutsos1994}.
Our 
real datasets come from the domains of seismology and astronomy. 
The seismic dataset, \emph{Seismic}, was obtained from the IRIS Seismic Data Access archive~\cite{url/data/seismic}. 
It contains seismic instrument recordings from thousands of stations worldwide and consists of 100 million data series of size 256,
i.e. its size is 100GB. 
The astronomy dataset, \emph{Astro}, represents celestial objects and was obtained from~\cite{journal/aa/soldi2014}.
The dataset consists of 270 million data series of size 256, i.e. its size is 265GB.
Since the main memory of our machine is limited to 256GB, we only use the first 200GB of the Astro dataset in our experiments.


\noindent{\bf Evaluation Measures.}
%
We measure (i) the {\em summarization time} required to calculate
the iSAX summaries and fill-in the summarization buffers, 
(ii) the {\em tree time} required to insert the items of the receive buffers in the tree-index,
and (iii) the {\em query answering time} required to answer 100 queries that are not part of the dataset. 
The  sum of the above times constitute the {\em total time}.
Experiments are repeated $5$ times and averages are reported.
All algorithms return exact results.

\subsection{Results}

\noindent{\bf \Fresh\ vs \MESSI.}
We compare \Fresh\ against \MESSI, which is the state-of-the-art blocking in-memory data series index. 
To enable a fair comparison, we use an 
optimized version of the original \MESSI\ implementation, 
where we have applied all the code 
enhancements incorporated by \Fresh.

We also compare \Fresh\ against an extended version of \MESSI, called \MESSIenh, that 
allows several threads to concurrently populate the same sub-tree, during tree creation
(instead of using a single thread for each subtree). 
This is implemented 
using fine-grained locks that are attached on each leaf node of a subtree.
\MESSIenh\ allows to compare the lock-free index creation phase of \Fresh\ against a more
efficient blocking one than that of original \MESSI.

Figure~\ref{fig:eval:fresh-messi-threads:seismic} shows that all algorithms (\Fresh, \MESSI, and \MESSIenh) 
continue scaling as the number of threads is increasing, for Seismic 100GB. 
This is true for all three phases. 
Moreover, the total execution time of \Fresh\ 
(Figure~\ref{fig:eval:fresh-messi-threads:random:total-from-4}
is almost the same as the total execution time of all its competitors, although it is the only lock-free approach. 
As expected, the tree index creation time of \Fresh\ is smaller than \MESSI's 
(Figure~\ref{fig:eval:fresh-messi-threads:random:tree-from-4}), since \Fresh\ allows subtrees to be
populated concurrently by multiple threads, allowing parallelism during this phase, in contrast
to \MESSI. 
Interestingly, \Fresh\ achieves better performance than \MESSIenh, in most cases.
The results for Seismic are similar and are omitted for brevity. 

Considering scalability as the size of the dataset increases, 
Figure~\ref{fig:eval:scale-dataset:seismic} demonstrates that \Fresh\ scales well on all three datasets. 
In most cases, \Fresh\ is faster than \MESSI.

Following previous works~\cite{DBLP:journals/vldb/ZoumpatianosLIP18,PFP21-I}, we also conducted experiments with query workloads of increasing difficulty.
For these workloads, we select series at random from the 
collection, add to each point Gaussian noise ($\mu = 0$, $sigma = 0.01-0.1$), and use these as our queries. 
Figure~\ref{fig:eval:scale-query-difficulty:seismic:total} presents the results for the Seismic dataset, where \Fresh\ performs better 
than \MESSI\ in most cases.

\begin{figure*}[h]
	\begin{subfigure}{0.24\textwidth}
		\includegraphics[width=0.94\textwidth]{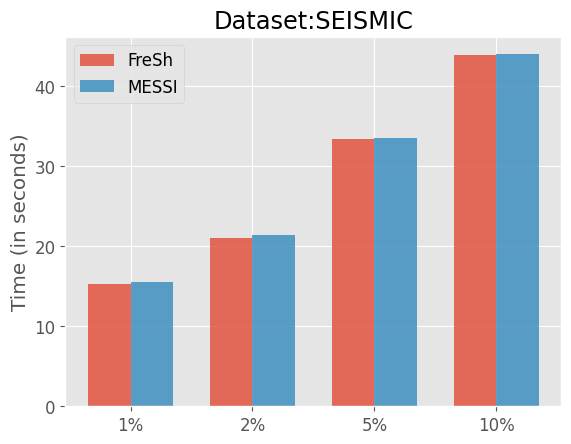}
		\caption{Total}
		\label{fig:eval:scale-query-difficulty:seismic:total}
	\end{subfigure}		
	\begin{subfigure}{0.24\textwidth}	
		\includegraphics[width=0.94\textwidth]{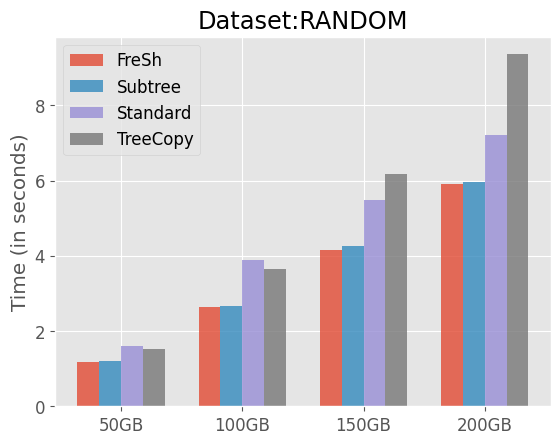}
		\caption{Tree Index}
		\label{fig:eval:scale-dataset:tree-index:random}
	\end{subfigure}		
	\begin{subfigure}{0.24\textwidth}	
		\includegraphics[width=0.98\textwidth]{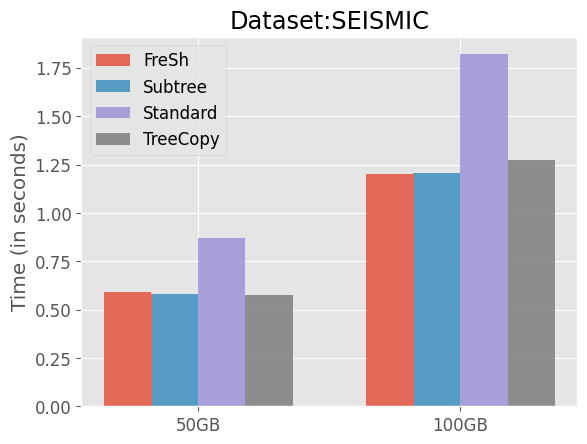}
		\caption{Tree Index}
		\label{fig:eva:scale-dataset:tree-index:seismic}
	\end{subfigure}		
	\begin{subfigure}{0.243\textwidth}	
		\includegraphics[width=0.98\textwidth]{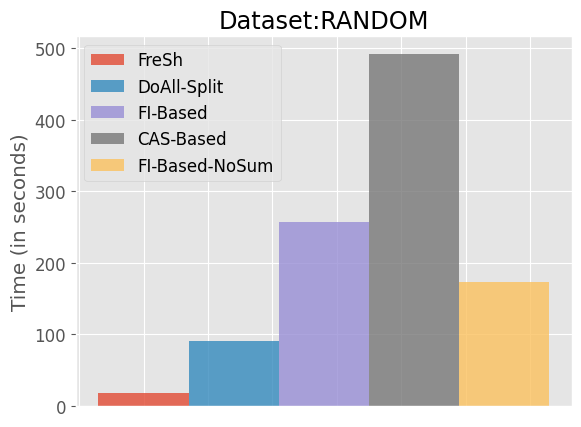}
		\caption{Total}
		\label{fig:eval:baselines:random:100GB:total}
	\end{subfigure}		
	\vspace*{-0.2cm}
	\caption{{\bf (a)} Comparison of \Fresh\ against \MESSI\ on Seismic 100GB with variable query difficulty, where an increasing percentage of noise is added to the original queries.
{\bf (b)-(c)} Comparison of \Fresh\ index creation to other tree implementations. {\bf (d)} Comparison of \Fresh\ against baseline implementations on 100GB Random.}
	\label{fig:eval:scale-dataset:tree-index}
	\label{fig:eval:scale-query-difficulty:seismic}
	\label{fig:eval:baselines:random:100GB}
\end{figure*}

\noindent
{\bf \Fresh\ vs Baselines.}
We compare \Fresh\ against several baseline {\em lock-free} implementations of the 
different stages of an iSAX-based index. 
Our results (Figure~\ref{fig:eval:baselines:random:100GB:total}
shows that \Fresh\ performs better 
than all these implementations.


\noindent
{\emph{\underline{Summarization Baseline:}}}
For buffer creation, 
we have experimented with three implementations:
\DoAllSplit, \FI, and \CASBased. 
All use a single summarization buffer with as many elements as \RawData.
%
%
\DoAllSplit\ 
splits \RawData\ into as many equaly-sized chunks as the number of threads.  
It stores a {\em done} flag with each data series, which is set after the data series is processed.
Each thread traverses \RawData\ (circularly), starting from the first element of its assigned chunk.
The thread first checks whether the done flag of a data series is set, and processes it only if not. 
%
%
In \FI, threads use \FAI\ to get assigned data series from \RawData\ to process.  
When a thread figures out that all \RawData\ elements have been assigned,
it re-traverses \RawData\ to identify data series whose done flag is still \False,
and processes them. 
\CASBased\ works similarly to \FI, while it uses \CAS\ instructions, instead of \FAI.
\Fresh\ performs significantly better than all these implementations
(Fig.~\ref{fig:eval:baselines:random:100GB:total}). 

\noindent
{\emph{\underline{Tree Population Baseline:}}}
Each thread is assigned elements of the summarization buffer using \FAI\
and inserts them in the index tree. 
To achieve lock-freedom in traversing the summarization buffer, 
we apply the \DoAllSplit, \FI, and \CASBased\ techniques we describe above.
%
To achieve lock-freedom in accessing the tree, we utilize a flagging technique~\cite{EFR+10},
in addition to our new tree implementation. 
%
We have also experimented with \FINoSum, a lock-free implementation that 
avoids using the summarization buffers and inserts directly iSAX summaries in the index tree, 
by applying the \FI\ technique on \RawData.
\Fresh\ performs significantly better than all these implementations
(graph omitted for brevity).

\noindent
{\emph{\underline{Pruning Baseline:}}}
All baselines use a single instance of an existing skip-based lock-free priority queue~\cite{LJ13}
to store the candidate data series for refinement.
Threads uses \FAI\ to find the next node to examine in the index tree.
%
When a thread $t$ discovers that all nodes 
of the tree have been assigned for processing, it re-traverses 
the tree to find nodes that may still  be
unprocessed, and processes them. 

\noindent
{\emph{\underline{Refinement Baseline:}}}
All threads, repeatedly call DeleteMin
to remove elements from the priority queue, and calculate their real distance computation. 
%
Our results (graph omitted due to lack of space) show that \Fresh\
performs significantly better than all these implementations, for query answering time (that includes pruning and refinement).

\noindent{\bf Performance breakdown for index creation phase.}
We evaluate the techniques incorporated by \Fresh\ to create its
tree index by comparing it against three modified versions of it. 
Recall that in \Fresh\ each thread populates each of the subtrees it
acquires in expeditive mode, as long as no helper reaches the same leaf of the tree;
when this happens it changes its execution mode to standard. 
So, \Fresh\ allows 
leaves of the same subtree to be processed in different modes of execution.

In the first modified version, called \FreshSub, threads start again by populating 
a subtree in expeditive mode, while they change to standard mode as long as a helper 
reaches this subtree (and not when it reaches one of its leaves, as \Fresh\ does); 
so, in \FreshSub\ all the leaves of a subtree are executed in a single mode at each 
point in time. 
In the second modified version, called \FreshSTD, threads populate subtrees using only 
the standard execution mode; i.e., there is no expeditive mode.
In the third modified implementation, called \FreshTreeCopy, a thread $t$ first populates a private
copy of the subtree (i.e. one that is accessible only to $t$) and only after its creation finishes,
$t$ tries to make it the (single) shared version of this subtree (by atomically changing a pointer 
using a \CAS\ instruction); threads help each other by following the same procedure.

Figures~\ref{fig:eval:scale-dataset:tree-index:random}-\ref{fig:eva:scale-dataset:tree-index:seismic}  compare \Fresh\ against the modified versions 
on Random and Seismic with variable dataset sizes and shows that it performs better than them,
in all cases. Interestingly, for Seismic 50GB \Fresh\ performs similarly to \FreshTreeCopy.
Recall that each thread works on its own private copy and, on each subtree, they contend at most once 
on the corresponding \CAS\ object. So, \FreshTreeCopy\ both restricts parallelism and minimizes the 
synchronization cost, which are properties that provide an advantage on Seismic. 

\noindent 
{\bf Thread Delays.} 
\remove{A  thread may be delayed due to page faults 
or more cache misses than other threads. Such delays may occur due to 
e.g. oversubscribing or time sharing. They may also 
occur in settings where the data series index may undergo long phases of
updates on the index. In lock-based solutiobs, qeries that need to access
specific parts of the tree index may be delayed beacuse other threads 
update them. Although this setting is not supported from any state-of-the-art
iSAX-based index, supporting updates is a highly desirable feature,
thus indicating a fruitful area of further research. Finally, threads
may experience failures due to software errors that may exist in the applications
that create them and utilize the data series indexes for answering queries. 
}
In order to study systems where processes may experience delays (e.g., due to page faults, time sharing, or long phases of updates),
we came up with a simplistic benchmark, where we simulate delays at random points of a thread's execution. 
%
Figure~\ref{fig:eval:failure-query:random:failure1} 
illustrates 
that the delay even of a single thread causes a linear overhead on the performance of \MESSI, 
whereas it hardly has any impact in the performance of \Fresh.
%
Moreover, Figure~\ref{fig:eval:failure-query:random:failure2} 
shows that \MESSI\ takes (almost) the full performance hit of delayed threads right from the beginning: 
even a single delayed thread blocks 
the execution of all other threads, and hence of the entire algorithm.
\Fresh\ gracefully adapts to the situation of increasing number of delayed threads, 
achieving a speedup. 
%
Note that these synthetic benchmarks are designed to simply illustrate the impact of lock-freedom 
on performance when threads may experience delays (or crash),
and not to capture some realistic setting. 
Note that \MESSI\ will not terminate execution even if a single thread fails. 
In the case of failures (see Figure~\ref{fig:eval:variable-num-failures}), \FreSh\ always terminates 
execution, performing almost identical to \MESSI\ with the same number of non-failing threads. 
This demonstrates that \FreSh\ adapts to dynamic thread environments, maintaining high performance levels.

\begin{figure}
	\begin{subfigure}{0.23\textwidth}
		\includegraphics[width=1\textwidth]{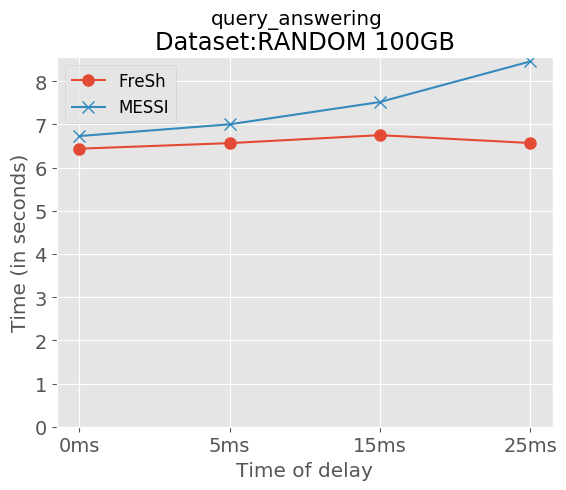}
		\caption{A single thread is delayed.}
		\label{fig:eval:failure-query:random:failure1}
	\end{subfigure}		
	\begin{subfigure}{0.23\textwidth}
		\includegraphics[width=1\textwidth]{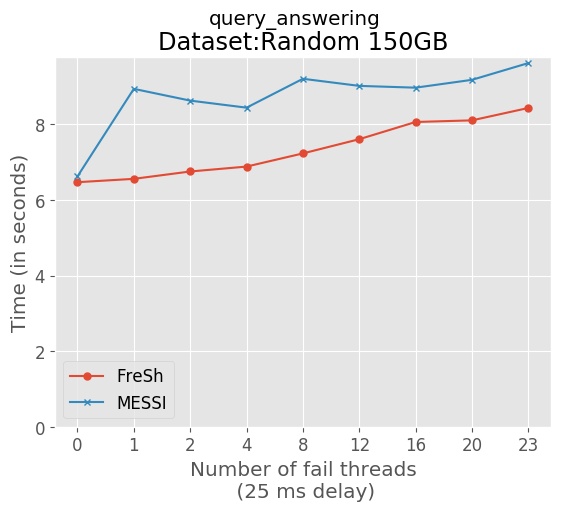}
		\caption{Mutliple threads are delayed. }
		\label{fig:eval:failure-query:random:failure2}
	\end{subfigure}		
	\vspace*{-0.2cm}
	\caption{Comparison of \Fresh\ against \MESSI\ 
		when varying delay and number of delayed threads.}
	\label{fig:eval:failure-query:random}
\end{figure}

\begin{figure}[t]
	\centering
	\begin{subfigure}[c]{0.24\textwidth}
		\includegraphics[width=1\textwidth]{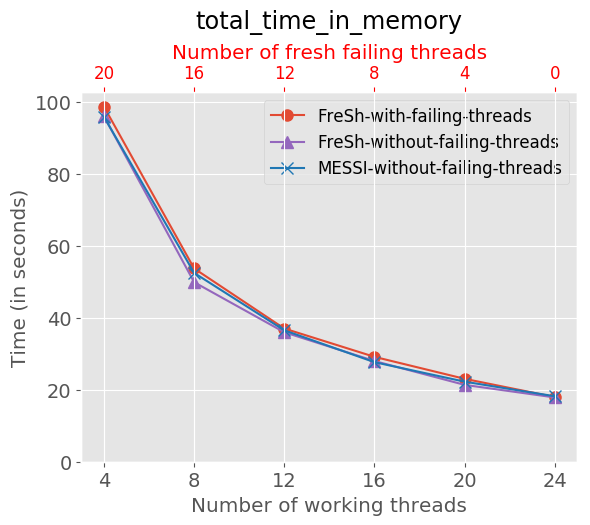}
		\caption{Total}
		\label{fig:eval:variable-num-failures:total}
	\end{subfigure}		
	\begin{subfigure}[c]{0.24\textwidth}	
		\includegraphics[width=1\textwidth]{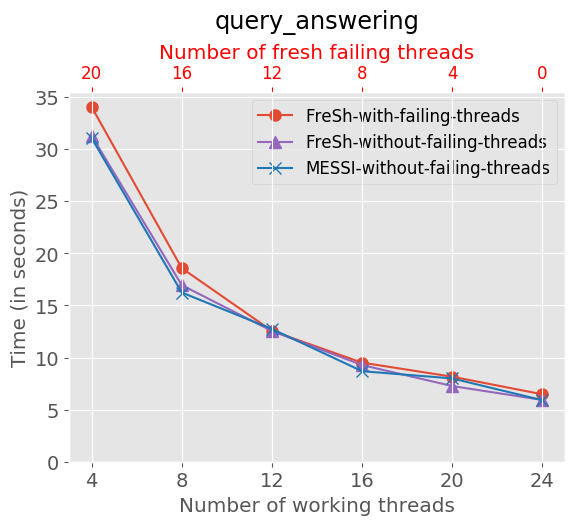}
		\caption{Query answering}
		\label{fig:eval:variable-num-failures:queries}
	\end{subfigure}		
	\vspace*{-0.2cm}
	\caption{Execution time on Random 100GB, when varying the number of threads that permanently fail 
	(\FreSh\ with permanent failures in red circles, \FreSh\ without failures in purple 
	triangles, \MESSI\ without failures in blue crosses).
	}
	\label{fig:eval:variable-num-failures}
\end{figure}

\remove{
\noindent{\em Single thread, multiple failures.} For each query, a single thread fails more than once in 
randomy chosen points during its execution. The number of times the thread fails is called {\em failure rate}.
We fix the failure delay at $2.5x$.
Figure~\ref{?} shows the results of this scenario while scaling the  failure frequency.
\lc{A variant of scenario 2 could be to fail with the same probability at each position, i.e. never select one of the 
positions at the beginning}.

\noindent{\em Many threads, multiple failures per thread.} For each query, each thread 
fails with failure rate $2$ and failure delay $2.5x$.
Figure~\ref{?} shows the results of this scenario while scaling the number of threads.

\noindent{\em Multiple threads, single failure per thread.}

\noindent{\em Oversubscribing.}

}

\section{Conclusions}


Current state-of-the-art data series indexes exploit the parallelism supported
by modern multicore machines, yet, their design is lock-based, and therefore, these implementations are blocking.
In this paper, we present \Fresh, a {\em lock-free} index, 
based on \Refresh, our novel generic approach for designing, building and analyzing highly-efficient data series indexes in a modular way. 
The experimental evaluation demonstrates that \Fresh performs as good as the state-of-the-art {\em blocking} index, thus, adhering to the same locality-aware design principles.

\vspace*{0.2cm}
\noindent{\bf [Acknowledgements]}
This work was supported by the Hellenic Foundation for Research and Innovation (HFRI) under the ``Second Call for HFRI Research 
Projects to support Faculty Members and Researchers'' (project number: 3684, project acronym: PERSIST). Part of the work of E. Kosmas was done while he was working at FORTH ICS.

%


\clearpage 
\bibliographystyle{IEEEtran}
\bibliography{references,odysseyref,edbttutorialref,hydraref}
\clearpage 
\remove{
\newpage
\appendix


\remove{

\begin{figure}[h!]
	\centering
	\begin{subfigure} [b]{0.13\textwidth}
		\centering
		\includegraphics[width=\textwidth]{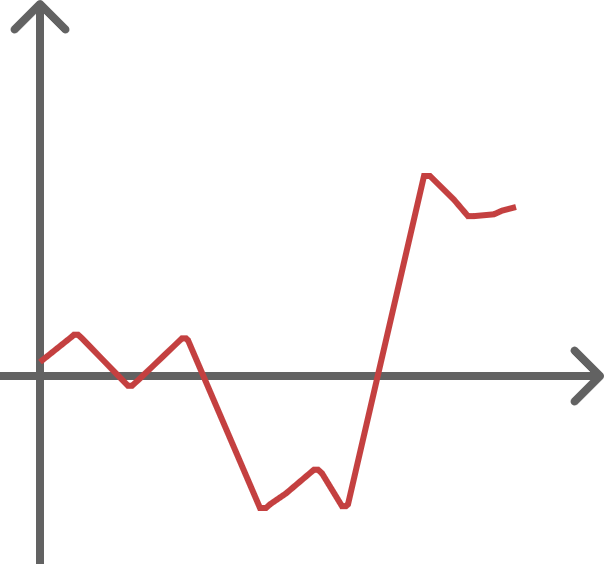}
		\caption{Data Series}
	\end{subfigure}
	\begin{subfigure} [b]{0.13\textwidth}
		\centering
		\includegraphics[width=\textwidth]{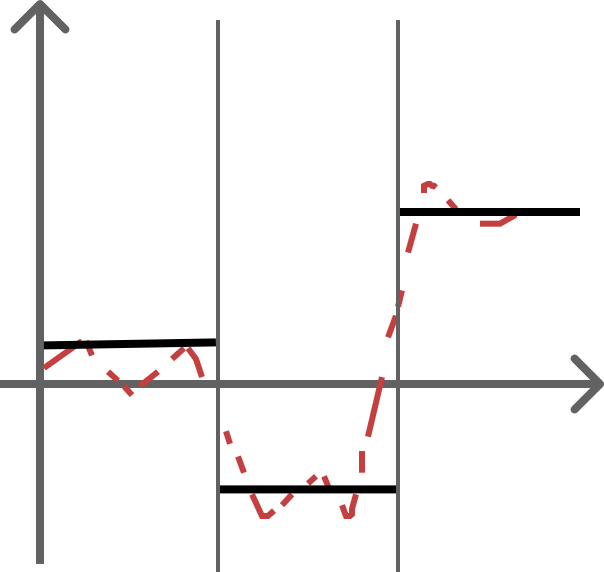}
		\caption{PAA Summary}
		\label{}
	\end{subfigure}
	\begin{subfigure} [b]{0.177\textwidth}
		\centering
		\includegraphics[width=\textwidth]{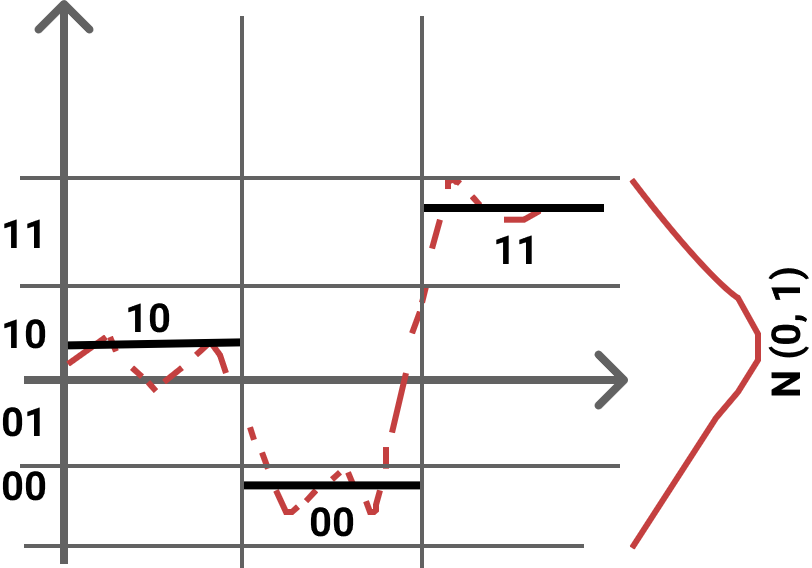}
		\caption{iSAX Summary}
		\label{}
	\end{subfigure}
	\begin{subfigure} [b]{0.38\textwidth}
		\centering
		\includegraphics[width=\textwidth]{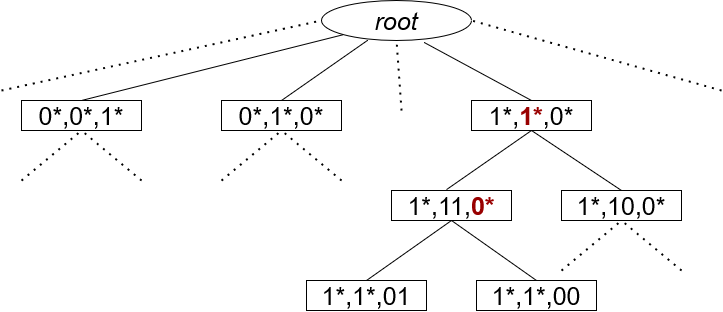}
		\caption{iSAX Tree}
		\label{}
	\end{subfigure}
	\vspace*{-0.2cm}
	\caption{From data series to iSAX index}
	\vspace*{-0.5cm}
	\label{fig:from_ds_to_iSAX}
\end{figure} 
}

\begin{figure}[h!]
    \centering
    \includegraphics[width=0.5\textwidth]{pdf/iSAX-index.pdf}
    \caption{Similarity search with the use of a data series index.}
    \label{fig:example}
\end{figure}

\begin{figure}[h!]
    \centering
    \includegraphics[width=0.5\textwidth]{pdf/flowchart2.pdf}
    \caption{Index building and query answering flowchart for the MESSI data series index.}
    \label{fig:example}
\end{figure}




\clearpage


\begin{algorithm}[t]
	\setcounter{AlgoLine}{0}
		\removelatexerror
		\footnotesize
		\begin{flushleft}	
			
			\com Shared variables:\;		
			\nl			TreeNode *$\mathit{IndexTree}[1..2^w]$ \label{alg:ps:r} \;
			\nl			\Boolean\ $\mathit{DTree[1..2^w]}$, $\mathit{HTree}[1..2^w]$, initially all \False \label{alg:ps:c}\;		
			\nl		CounterObject $\mathit{TreeCnt[1..2^w]}$\;
		\end{flushleft}
		
		\vspace*{1mm}
		
		\myproc{{{\small \Traverse}(Function *\Prunning, TreeNode *$T$, CounterObject *$\mathit{Cnt}$, int $x$, Boolean $h$, int $\mathit{rlevel}$)}}{
			\nl		int $\mathit{i}$\;
			
			\nl		\While{$\True$}{
				\nl			$\mathit{\langle i, * \rangle = Cnt.\NextIndex(\&h)}$		\label{alg:ps:help:c:cnt}	\;
				\nl			\lIf{$\mathit{i > x}$}{\Break}
				\nl			\uIf{$\mathit{rlevel < 2}$}{ 
					\nl				{\bf mark} $\mathit{IndexTree[i]}$ as acquired \;
					\nl				$\mathit{totalNds}$ := \TotalNodes($\mathit{IndexTree[i]}$)\;
					\nl	   			\Traverse($\mathit{Prunning, IndexTree[i], TreeCnt[i]}$, \\ 
					\hspace*{1cm} $\mathit{totalNds, \False, rlevel+1}$)			\label{alg:ps:help:c:rec}\;
					\nl				$\mathit{DTree[i] := \True}$						\label{alg:ps:help:c:true}
				}
				\nl			\uElse{ \label{alg:ps:em:else}
					\nl				$\mathit{nd}$ := \FindNode($\mathit{T,i}$)		\label{alg:ps:findnode} \;
					\nl				{\bf mark} $\mathit{nd}$ as acquired \;
					\nl				*\Prunning($\mathit{nd}$)				\label{alg:ps:prunning}\;	
					\nl				{\bf mark} $\mathit{nd}$ as done \;
				}
			}
			\nl	 	\For{\Each $j$ such that $\mathit{DTree[j]}$ is equal to \False} { \label{alg:ps:scan:for} 
				\nl			\Backoff()	\tcp*{avoid helping, if possible}				\label{alg:ps:help:backoff}
				\nl			\uIf{$\mathit{DTree[j]] = \False}$}{						\label{alg:ps:help:if}
					\nl				$\mathit{HTree[j] := \True}$ \;						\label{alg:ps:h:true}
					\nl					\HelpTree(\Prunning, $\mathit{IndexTree[j]}$)\;
				}
				\nl				$\mathit{DTree[j] := \True}$						\label{alg:ps:help:c:true}
			} 
		}
		\vspace*{1mm}	
		\myproc{{{\small \FindNode}(TreeNode *$T$, int $i$): returns TreeNode*}}{			\label{alg:ps:help:c:treenode-start}
			\nl		TreeNode *$\mathit{p = T}$\; 
			\nl		int $\mathit{nds = 0}$\;
			
			
			\nl		\While{$\mathit{p \neq NULL\ \And\ nds \neq i}$}{
				\nl			\uIf{$\mathit{nds + p \rightarrow cnt + 1 < i}$}{
					\nl				$\mathit{nds = nds + p \rightarrow cnt + 1}$\;	
					\nl				$\mathit{p = p \rightarrow rc}$\;
				}
				\nl			\lElse{
					$\mathit{p = p \rightarrow lc}$
				}
			}
			\nl		\Return $\mathit{p}$\;							\label{alg:ps:help:c:treenode-end}
		}
		\vspace*{1mm}	
		\myproc{{{\small \HelpTree}(Function *$f$, TreeNode *$T$)}}{
			\nl		\lIf{T == NULL} { \Return }
			\nl		\HelpTree($f$, $T \rightarrow lc$)\; 
			\nl		\lIf{$*T$ is unprocessed}{ *$f$($*T$)}
			\nl		\HelpTree($f$, $T \rightarrow rc$)\;
		}
		\caption{Pseudocode for \Put\ and \Traverse\ of \PS\ in \Fresh. Code for thread $t \in \{ 1, \ldots, N-1\}$.}
		\label{alg:ps}
	\end{algorithm}

\input{tree-code}

\input{pq-code}

\newpage

}
\end{document}